\documentclass[preprint]{aastex}
\usepackage{emulateapj5}
\def\stacksymbols #1#2#3#4{\def\theguybelow{#2}
	\def\verticalposition{\lower#3pt}
	\def\spacingwithinsymbol{\baselineskip0pt\lineskip#4pt}
	\mathrel{\mathpalette\intermediary#1}}
\def\intermediary #1#2{\verticalposition\vbox{\spacingwithinsymbol
	\everycr={}\tabskip0pt
	\halign{$\mathsurround0pt#1\hfil##\hfil$\crcr#2\crcr
		\theguybelow\crcr}}}
\def\lta{\stacksymbols{<}{\sim}{2.5}{.2}}
\def\gta{\stacksymbols{>}{\sim}{3}{.5}}

\begin{document}

\title{ENTROPY EVOLUTION IN GALAXY GROUPS AND CLUSTERS: 
A COMPARISON OF EXTERNAL AND INTERNAL HEATING}

\author{Fabrizio Brighenti$^{1,2}$ and William G. Mathews$^1$}

\affil{$^1$University of California Observatories/Lick Observatory,
Board of Studies in Astronomy and Astrophysics,
University of California, Santa Cruz, CA 95064\\
mathews@lick.ucsc.edu}

\affil{$^2$Dipartimento di Astronomia,
Universit\`a di Bologna,
via Ranzani 1,
Bologna 40127, Italy\\
brighenti@bo.astro.it}

%\vskip 2.in
%\noindent
%Received:

%\noindent
%PROOFS TO BE SENT TO:

%\noindent
%Lick Observatory

%\noindent
%Santa Cruz, CA 95064

%\noindent
%$^1$UCO/Lick Observatory Bulletin No.

%\vskip3.in
%\noindent
%Short Title: X-ray Images of Elliptical Galaxies
%\clearpage
\vskip .2in

\begin{abstract}

The entropy in hot, X-ray emitting 
gas in galaxy groups and clusters 
is a measure of past heating events, 
except for the entropy lost 
by radiation from denser regions. 
Observations of galaxy groups indicate higher entropies 
than can be achieved in the accretion shock experienced 
by gas when it fell into the dark halos. 
These observations generally refer to the dense,
most luminous inner regions where the 
gas that first entered the halo may still reside. 
It has been proposed 
that this non-gravitational entropy excess 
results from some heating process in the 
early universe which is external to the group and 
cluster halos and that it occurred 
before most of the gas had entered the dark halos. 
This universal heating of cosmic gas 
could be due to AGN, population III stars, 
or some as yet unidentified source.
Alternatively, the heating of the hot gas 
in groups may be produced internally by Type II 
supernovae when the galactic stars in these systems formed.
We investigate here the consequences of 
various amounts of external, high redshift 
heating with a suite of gas dynamical calculations. 
We consider the influence of radiation losses and distributed 
mass dropout on the X-ray 
luminosity and emission-weighted 
temperature of the hot gas as well as its 
central entropy. 
In general we find that 
externally heated flows are unsatisfactory; 
when the heating is high enough to bring the 
X-ray luminosities into agreement with observations, 
the gas entropy is too high.
We compare these solutions with flows that are internally 
heated by Type II supernova; this type of heating depends 
on the IMF and the efficiency that the supernova energy 
is conveyed to the hot gas. 
These internally heated flows give much better agreement 
with X-ray observations of galaxy groups 
and are insensitive to the levels of supernova heating 
that we consider as well as to the epoch and 
spatial distribution of the supernova heating process.
However, to fit X-ray observations 
a large fraction of the energy produced by high redshift
Type II supernovae must heat the hot gas 
if the number of supernovae is based on 
a Salpeter initial mass function.
Alternatively, only about 20 percent of the Type II 
supernova energy would be required to heat the gas if 
the initial mass function has a flatter slope 
than Salpeter, as suggested by stellar mass to light ratios.
\end{abstract}


%\clearpage

\section{INTRODUCTION}

In a perfect, starless $\Lambda$CDM hierarchical universe
filled with adiabatic gas and
NFW (Navarro, Frenk, \& White 1996) dark halos,
the bolometric 
X-ray bremsstrahlung luminosities of galaxy groups and
clusters would scale in a self-similar
fashion with gas temperature,
$L \propto T^2$ (Kaiser 1986; 1991; Evrard \& Henry 1991).
However, in our particular universe this relation is somewhat
steeper ($L \propto T^3$; e.g. Arnaud \& Evrard 1999)
and becomes very steep
(at least $L \propto T^4$) for groups having
$T \lta 1$keV (Helsdon \& Ponman 2000a).
The observed properties of galaxy groups differ from
those of massive clusters in several other respects.
Groups have a lower baryon fraction
$f_B = \Omega_b/\Omega_o$ 
(David 1997; Renzini 1997)
and a larger fraction of group baryons are stellar
(e.g. David et al. 1990; David \& Blumenthal 1992).
Finally, the entropy factor $S \equiv T/n_e^{2/3}$
evaluated at $0.1r_{vir}$ for groups
exceeds the self-similar expectation,
indicating that groups have received
an additional entropy $S \sim 100$ keV cm$^2$.
If a comparable entropy increment
were present in the hot gas in massive clusters, it would be difficult
to detect because of the much larger entropy that the gas acquires
passing through stronger accretion shocks that surround more massive
clusters.
This has been referred to as the ``entropy floor''
(Ponman et al. 1999; Lloyd-Davies et al. 2000).

These deviations from self-similarity
has led to the hypothesis that
gas in groups experienced some additional early heating
before (or when) it flowed into the group halos.
Early ``pre-heating'' by 0.5-1.5 keV/particle could explain the
aberrant behavior of groups in both the $L-T$ and
$S-T$ plots.
Furthermore, since this level of heating would only
be noticed in the shallower potentials of galaxy groups,
most authors have adopted a much stronger
assumption: that {\it all} baryonic gas, including gas
currently in both groups and clusters,
experienced some ``non-gravitational'' heating
(star formation, Pop III stars,
AGN, etc.)
prior to its entry into the dark halos.

As a result of its dissipative nature, the gas entropy observed
in galaxy groups
today does not in itself indicate a unique heating history.
Entropy increases in shocks and is lost with
radiative cooling.
Nevertheless,
the level of heating required to heat all the gas 
at the same (high) redshift in
``pre-heating'' scenarios generally exceeds that
produced by Type II supernovae following normal star formation:
$\sim 0.2$ keV/particle.
In this estimate we assume 
ten percent of the baryons form into stars 
(Fukugita, Hogan \& Peebles 1998)
with a typical Salpeter IMF, 
producing Type II supernovae that heat the 
remaining gas with 100 percent efficiency. 

Nevertheless, many theoretical papers have appeared recently
with estimates of the assumed universal
pre-heating necessary to produce the ``entropy floor'' in
groups and the related departure from $L - T$ self-similarity
(e.g. Knight \& Ponman 1997;
Cavaliere, Menci, \& Tozzi 1997;
Balogh, Babul \& Patton 1999;
Cavaliere, Giacconi \& Menci 2000;
Loewenstein 2000; Tozzi \& Norman 2000;
Valageas \& Silk 1999; Kravtsov \& Yepes 2000).
For a given energy release, 
the final entropy is larger if the energy is applied
when the gas density is low.
Therefore,
most of these authors have assumed that the gas
was heated (by some unspecified agency) 
while in a low density intergalactic
environment (at redshift $z \lta 7$)
before it flowed into the galaxy group potentials.

Independent support for strong universal heating has come from
estimates of the collective emission from 
gas-filled group halos at large redshifts
which appear to exceed the 
observed soft X-ray background radiation
(Pen 1999; Wu, Fabian \& Nulsen 1999b or WFN).
These authors suggest that the
intergalactic gas was heated to $T \gta T_{vir}$
so that little of it flowed into group halos
at early times.
However, SNII-driven galactic winds are expected to develop
immediately in these small halos, 
perhaps reducing the contribution of groups
to the unresolved X-ray background.
In addition, warm-hot diffuse intergalactic
gas at temperatures $10^5 \lta T \lta 10^7$K
can result from (gravitationally produced)
shocks in large-scale filaments (Cen \& Ostriker 1999;
Dav\'{e} et al. 2000). 
Moreover, 
in the cosmological simulations of Dav\'{e} et al. 
most of the warm-hot gas 
that contributes to the unresolved soft X-ray background 
is at low densities and 
diffusely distributed, not concentrated 
in (group) halos as assumed by Pen and WFN.
Consequently, Dav\'{e} et al. predict a 
soft X-ray flux $\sim 100$ times less than that 
of Pen and WFN, sufficiently low to be consistent with 
the observed (unresolved) background. 
Although the non-gravitational heating of intergalactic
gas postulated by Pen and WFN may occur, 
it is not supported by more detailed 
cosmological simulations.

Tozzi \& Norman (2000) study the consequences
of universal 
intergalactic heating prior to collapse into dark halos
in a flat $\Lambda$CDM cosmology.
In their approximate hydrodynamic models
with radiative cooling,
all cooled baryonic gas is assumed to collect at the
very center.
The universal energy input (at $z > 1$) they require to match the
observed ``entropy floor'' is in excess
of normal SNII expectations.
In their preferred models, Tozzi \& Norman
assume that the intergalactic gas
(heated to $\sim T_{vir}$ for groups) enters the group halos
adiabatically, i.e. without shocking.
As a
consequence, the radial entropy profile for the Tozzi-Norman
externally pre-heated galaxy groups is almost constant out to the
virial radius $r_{vir}$, unlike those observed by Lloyd-Davies, Ponman
\& Cannon (1999) for which $S \propto r$, and the gas temperature
gradients of these Tozzi-Norman models are quite negative throughout,
e.g. $d \log T / d \log r \approx -0.4$ at $r \approx 0.1r_{vir}$,
unlike the nearly constant temperature profiles observed.
Both of these key results of adiabatic
inflow are in conflict with observed groups, 
at least within the small region that can be observed,
$r \lta (0.1 - 0.4)r_{vir}$.
By contrast, an internal heating scenario
is proposed by Loewenstein (2000) who estimates the
amount of heating required after the hot gas
has reached hydrostatic equilibrium in
group and cluster dark halos.
From a series of approximate static hot gas models,
Loewenstein argues that most of the heating occurred during
or after the assembly of the group or cluster gas, not
throughout the intergalactic gas at an earlier time. 
We are in agreement with his interpretation, 
although it may be a 
minority opinion at the present time.

Perhaps
the most convincing evidence that the origin and evolution of
hot gas in
galaxy groups can be understood with normal star formation
and other standard astrophysical
assumptions is the success of our own
detailed calculations for
the giant elliptical NGC 4472 (Brighenti \& Mathews 1999a),
the dominant galaxy in a small Virgo subcluster.
This is a complete gas dynamical calculation 
beginning with a tophat perturbation in a flat cosmology.
The gas evolves in a growing NFW dark halo, 
produces stars and SNII heating, forms a central galaxy 
and the stars lose mass and produce Type Ia supernova.
After 13 Gyrs we were able to match the
radial variation of density, temperature and iron abundance
observed in NGC 4472 (Brighenti \& Mathews 1999b).
The agreement is excellent everywhere except
within $\sim 1$ kpc from the center where
the gas density
is too high; we now think this is due to additional
support by magnetic stresses there.
Our models also agree
with the present day entropy variation and luminosity
in NGC 4472.
In particular, our models for NGC 4472
produce the ``entropy floor''
in the $S - T$ plot and the observed deviation
from the $L - T$ self-similar relation
without universal pre-heating.
In addition, they go much further in fitting the 
radial profiles of density, temperature and 
iron abundance (Buote 2000b) observed in NGC 4472.

The pre-heating controversy devolves on a
choice between {\it internal} supernova heating during
galaxy group formation and {\it external} (universal)
heating by AGN or some other agency at an early time.
Given the apparent success of our calculations
for NGC 4472, it seems possible that
the ``entropy floor'' and $L - T$ similarity breaking
in groups can be explained with normal star formation.
But the widely discussed pre-heating hypothesis 
can under some conditions also provide the 
necessary similarity-breaking.
Our objective in this paper is to perform 
a series of gas dynamical 
calculations that explore the consequences 
of various levels of external pre-heating 
that span the full range from galaxy groups to 
rich clusters. 
Several sets of progressively more sophisticated 
calculations are considered, 
beginning with purely adiabatic flow.
These externally heated cooling flows 
are compared with flows that only 
experience internal supernova-based heating.
We shall find that the latter mode of heating is generally 
quite satisfactory in fitting the X-ray data, 
but the IMF may need to be flatter than Salpeter.

\section{OVERVIEW OF 
GAS DYNAMICS OF GALAXY CLUSTERS AND GROUPS}

In order to compare our gas dynamical 
calculations with internal supernova heating 
with discussions of external 
pre-heating in the current literature,
we consider gas flowing into dark halos 
of three masses representing galaxy groups, 
poor clusters and rich clusters.
The dark halo evolution is similar in all models.
These structures all begin with a tophat perturbation 
in a $\Lambda$CDM universe
($\Omega_o = 0.3$, $\Lambda = 0.7$, 
$h = H_o/(100~{\rm km/s~Mpc}) = 0.725$) 
which converts to a dark matter NFW profile 
(Navarro, Frenk \& White 1996) at very high redshifts.
The dark halos are assumed to grow masswise from 
the inside out as described by 
Bertschinger (1985).
The dark matter flow pattern consists of 
an outer, converging smooth collisionless flow 
which attaches to a stationary NFW core 
having a concentration appropriate to its 
current virial mass.
The time-dependent 
intersection radius of the stationary halo with the 
converging flow is chosen to conserve  
dark matter mass.

Baryonic gas with mean density
$\Omega_b = 0.039 (h/0.7)^{-2}$ 
(Burles \& Tytler 1998a,b) 
flows into the evolving 
dark matter potential. 
However, in order to explore the cosmic 
pre-heating hypothesis using the same approach 
as often employed in the current literature,
we may choose to ``reset'' the 
density and temperature (and therefore the 
entropy) of the baryonic gas 
to be spatially uniform at a very early time 
$t_h = 0.5$ Gyr (redshift $z_h = 9$) 
when the stationary dark 
halo mass is only 0.1 - 0.2 of its current virial mass. 
At the reset, the gas velocity is assumed to 
retain its velocity in the cosmic flow  
outside the NFW halo at time $t_h$, 
but within the halo the gas velocity is 
set to zero.
This artificial reset of the gas parameters 
at this high redshift often has no long-lasting 
influence on the gas dynamics, however, 
if the gas temperature is less than the virial 
temperature of the dark matter structures 
at time $t_h$, shocks may develop resulting in 
entropy fluctuations that persist until the 
present time $t_n = 13$ Gyrs.

In the following discussion
we consider a suite of increasingly more realistic 
physical models for the baryonic component.
For dark halos of each mass, the baryonic 
gas is assumed to be pre-heated at time 
$t_h = 0.5$ Gyr by varying amounts.
For the first set of models, 
we assume that the gas flow is perfectly
adiabatic but with adiabatic (non-radiating) shocks, 
so most of the entropy increase 
occurs in the accretion shock transition. 
In the next series of models we 
include radiative cooling and allow the 
cooled gas to accumulate at the origin 
as in the models of Tozzi \& Norman (2000).
Next we repeat some of these models with 
a cooling recipe similar to that used 
by Nulsen \& Fabian (1997) in which the cooled 
gas accumulates not at the origin, but 
in spatially extended structures, 
as if it had formed into a collisionless 
stellar system. 
This type of flow is similar to the 
mass dropout flows discussed 
by Knight \& Ponman (1997) 
(but who assumed Bertschinger (1985) dark 
halos in a $\Omega_o = 1$ cosmology with a 
large baryon fraction and no heating).
For each series of calculations 
we consider a range of virial masses
and assumed entropies corresponding to 
different levels of external pre-heating
at the reset time.
Finally, we repeat the same calculations again 
allowing for star formation with various
amounts of energy released
in Type II supernova (SNII) feedback.
In this series of models, which resembles 
our earlier calculation for NGC 4472, 
the SNII heating at early times is internal, 
a natural consequence of star formation. 
These models also include stellar mass loss 
and heating by Type Ia supernovae. 

\subsection{Observational Data}

The X-ray luminosity varies over a factor 
$\gta 10^4$ from galaxy groups to the richest clusters
and the gas temperature spans a range 
of $\sim 30$.
Because of the inhomogeneous observational data available 
over this vast range of parameters 
and the various bandpasses that have 
been used in these observations, 
we shall consider only those observations 
for which the bolometric X-ray luminosities 
are determined.
All data is corrected to our assumed 
$H_o = 100 h = 72.5$ km s$^{-1}$ Mpc$^{-1}$.

For galaxy group luminosities and temperatures 
we choose the 
recent observations of Helsdon \& Ponman (2000a)
for loose groups. 
More recently, these authors 
have shown that there is no distinction in the 
X-ray properties of loose and compact 
groups (Helsdon \& Ponman 2000b).
Helsdon \& Ponman list ROSAT PSPC bolometric 
luminosities for 24 groups. 
Unfortunately, these groups are observed out to a radius 
which is typically only $\sim 0.1 - 0.2$ of the 
virial radius, 
so Helsdon \& Ponman provide for each group an approximate 
correction factor to estimate the  
total bolometric luminosity 
within the virial radius; we have 
employed this correction in their group data  
plotted with open squares in Figure 1.
The $L - T$ variation is quite steep 
(and uncertain) for 
these groups, $L \propto T^{4.9 \pm 0.8}$.
For the entropy in galaxy groups and clusters 
we use the data 
of Lloyd-Davies, Ponman \& Cannon (2000) who estimate 
the entropy factor $S = T n_e^{-2/3}$ at $0.1r_{vir}$.

For a sample of poor and rich clusters we use data for  
24 clusters collected by Arnaud \& Evrard (1999) from 
many disparate authors using different 
X-ray satellites.
In selecting this sample 
Arnaud \& Evrard avoided clusters with 
strong cooling flows.
The Arnaud-Evrard bolometric luminosities have not 
been corrected to the virial radius, 
but this correction may not be large. 
For example, the mean correction found by 
Helsdon \& Ponman for the groups is only 
a factor of $\sim 1.5 \sim 10^{0.2}$.
The Arnaud-Evrard data, shown with crosses 
in Figure 1, can be fit with 
$L \propto T^{2.88 \pm 0.15}$, 
which is also steeper than adiabatic similarity 
$L \propto T^2$. 

Allen \& Fabian (1998) emphasize the important 
effect that a strong cooling flow has on the 
$L - T$ relation.
About 70 - 90 percent of rich clusters have 
cooling flow cores.
Since the gas temperature has a steeper  
radial gradient (and is generally lower) in 
clusters with strong cooling cores, 
it is unclear which temperature 
is appropriate to enter in the $L - T$ diagram.
If the ASCA X-ray spectra of Allen \& Fabian 
are fit with isothermal models, 
(their Model A) the cooling flow clusters 
lie at lower temperatures 
(and higher luminosities) 
than non-cooling flow clusters 
as shown in Figure 1. 
The non-cooling flow clusters have 
slope $L \propto T^{2.90 \pm 0.29}$, 
very similar to the Arnaud-Evrard slope, 
and in fact appear to be fully consistent 
with the overlapping Arnaud-Evrard data.
Alternatively,  
if the X-ray spectra of cooling flow clusters 
are fit with a combination of single temperature 
thermal emission and XSPEC cooling flow models 
(in which the emitted spectrum is an integrated 
sum over many gas temperatures)
as in Model C of Allen \& Fabian (1998),  
then the background (uncooled) gas 
temperatures of cooling flow clusters 
increase and become very similar to those of the
non-cooling flow clusters in the Allen-Fabian sample.
This is a clear example of the substantial 
scatter that can occur in the $L - T$ diagram 
arising from the data reduction procedure 
when a wide range of temperatures contribute 
to the observed spectrum.
In a related approach, Markevitch 
(1998) showed that the $L - T$ correlation 
tightened considerably when the cooling flow cores 
are removed from both the luminosity and temperature 
data. 

When data for all groups and clusters 
are compared in Figure 1, 
it can be seen that 
the Arnaud-Evrard data have the same 
slope and ordinate as the non-cooling 
flow Allen-Fabian clusters.
This is expected since the Arnaud-Evrard 
clusters were also chosen to have weak 
cooling flow cores.
If Arnaud \& Evrard had instead chosen 
clusters at random, 
they would have found a much larger number of 
normal cooling flow clusters 
scattered above their 
correlation in Figure 1, similar to the 
cooling flow clusters in the Allen \& Fabian 
sample.
The identification of the Arnaud-Evrard 
clusters with 
the lower envelope of the correlation in 
the $L - T$ plot also explains why 
the Helsdon-Ponman poor groups 
with the greatest temperatures may 
seem slightly overluminous relative to adjacent 
members of the Arnaud-Evrard sample 
having the lowest temperatures.
This lack of consistency 
and apparent continuity between 
the various data sets demonstrates how 
systematic shifts in the $L - T$ plot 
may arise from different methods of 
sample selection and data reduction 
for the majority of cooling flows 
in which the intrinsic temperature 
necessarily varies over a wide range.

Since the $L - T$ data for groups are so important 
in setting limits on the amount of cosmic pre-heating,
some possible systematic trends should be 
recognized.
Just as with rich clusters, groups with strong cooling 
flows tend to occupy the upper envelope of the 
group data in Figure 1.
To demonstrate this we use the sample of 12 galaxy groups 
studied by Buote (2000a) in which the data was 
reduced with an XSPEC cooling flow model 
with a determination of the cooling rate 
${\dot M}$.
If the Buote data is bisected with a line of slope 
$L \propto T^3$ or $T^4$, the mean cooling flow rates of 
groups that lie above the line, 
$\langle {\dot M} \rangle = 21$ $M_{\odot}$ yr$^{-1}$, 
is much greater than the mean of the six groups below,
$\langle {\dot M} \rangle = 6.2$ $M_{\odot}$ yr$^{-1}$. 
X-ray emission from the more luminous sample 
have spuriously lower temperatures due to 
strong cooling flows associated with the central galaxy.
This is an important distinction for our models 
since we are interested in groups in which 
a strong cooling flow develops around a dominant,
luminous elliptical galaxy located at the group center.
For this reason we expect our successful models 
with concentrated cooling flows  
to lie near the upper envelope of the Helsdon-Ponman 
group data.
Additional scatter in the $L - T$ data for groups 
is introduced by the 
variable physical extents of the X-ray observations 
relative to the virial radius for each group. 

Finally, the choice of cosmology has some influence 
on the observational group data.
In extrapolating $L$ to the virial radius, 
Helsdon \& Ponman (2000a) used the 
SCDM adiabatic results of Navarro, Frenk, \& White (1995) to 
derive $r_{vir}$ from the observed gas temperature.
Our adiabatic models and $\Lambda$CDM 
variables give 
$r_{vir} \approx 2.3 (T/5.1{\rm keV})^{1/2}$ Mpc
which is only about 10 percent smaller than 
that used by Helsdon \& Ponman, 
probably within the observational and 
computational uncertainties. 

\subsection{Growth of NFW Profiles}

We consider three virial masses, 
$M_{vir} = 4.7 \times 10^{13}$, $2.2 \times 10^{14}$ 
and $1.2 \times 10^{15}$ $M_{\odot}$ which span a range 
from galaxy groups to moderately rich clusters.
These dark halos form as a result of a 
tophat perturbation in a flat $\Lambda$CDM 
cosmology, $\Omega_o = 0.3$ $\Lambda = 0.7$ and 
$H_o = 72.5$ km s$^{-1}$ Mpc$^{-1}$, 
chosen so that the current time is exactly 
$t_n = 13$ Gyrs. 
The baryonic component 
$\Omega_b = 0.039 (h/0.7)^{-2}$ is such that 
$\Omega_b/\Omega_o = 0.123$. 
When a tophat perturbation is imposed, 
the dark matter flows toward the perturbation 
as a pressure-free gas and accumulates in  
an extended, static, spherically symmetric structure.
This structure is assumed to 
have the standard NFW 
(Navarro, Frenk \& White 1996) halo profile 
for which the virial mass is 
$M_{vir} = (4 \pi / 3) r_{vir}^3 \Delta_c \rho_c$. 
Here $\Delta_c = 178 \Omega_o^{0.45} \approx 100$ 
(Eke, Navarro \& Frenk 1998) 
and $\rho_c = 3 H_o^2/8 \pi G 
= 1.07 \times 10^{-29}$ gm cm$^{-3}$ 
is the critical density.
The shape of the dark halo at any time is 
$M(r) = M_{vir} f(y)/f(c)$
where $f(y) = ln(1+y) - y/(1+y)$, 
$y = r/r_s = cr/r_{vir}$ and $c = r_{vir}/r_s$
is the concentration. 
Cosmological simulations show that the concentration 
decreases slowly with $M_{vir}$ and redshift
(Bullock et al. 2000).
We ignore the redshift 
variation, i.e., 
$c = 8.35 (M_{vir} h^{-1}/10^{14} M_{\odot})^{-0.0911}$.
The virial radii for the three dark halos are 
$r_{vir} = 895$, 1500, 2640 kpc or
$\log r_{vir} = 2.95, 3.17$ and 3.42 respectively.

The complete
solution for the dynamical growth of a locally overdense region of
collisionless dark matter in a flat Einstein-de Sitter universe was
solved by Bertschinger (1985).  He showed that the converging
pressure-free flow accumulates in an essentially stationary core which
grows masswise from the inside outward, as if newly arriving matter
were added at the outer boundary of the core.  Since Bertschinger made
no allowance for the hierarchical, three dimensional nature of mass
accumulation, the mass profile in his stationary core differs from the
more appropriate NFW shape.  We adapt the basic physics of
Bertschinger's collisionless solution for the evolution of a dark
matter overdensity to our flat $\Lambda$CDM cosmology by computing the
pressure-free collapse of the dark matter until it accumulates at the
outer boundary of a stationary halo that is required to have the more
correct NFW profile (Brighenti \& Mathews 1999b).
The halo grows with time (from the inside 
out) at a cosmologically 
appropriate rate, conserving dark mass. 
However, at very high redshifts 
($z > z_{tr} \sim 15$) this simple recipe 
breaks down for large $M_{vir}$ because the internal mass 
distribution of the NFW halo 
can be less than that of the converging $\Lambda$CDM 
flow at every radius.
When this mismatch occurs at high redshift, 
we assume that the dark matter 
collects into an extended, mass-conserving power law halo 
$M(r) \propto r^2$ within 40 kpc, 
which resembles the inner profile of a NFW halo.
This transition in the dark matter halo is 
particularly 
necessary for the most massive halo we consider,
NFW halos of smaller mass intersect the 
$\Lambda$CDM flow at very high redshifts 
($z_{tr} \gg 15$).
We find that the exact shape of the temporary, 
non-NFW dark halo for $z > z_{tr}$ 
has a only a small 
influence on the integrated properties of 
the gas at time $t_n = 13$ Gyrs. 
While none of our conclusions are affected by 
assumed dark halo structure at $z > z_{tr}$,
it is good to keep in mind that some of 
our computed results (e.g. final $L$, $T$, etc.) 
are uncertain at the $\sim 20$ percent level.
Finally, we note that the baryonic gas does 
not exactly follow the dark matter in the 
external $\Lambda$CDM flow since in general 
the gas velocity 
is slowed by radial pressure gradients.

\subsection{Gas Dynamics}

The equations for one dimensional 
Eulerian gas dynamics 
that we use here are described in detail in 
Brighenti \& Mathews (1999a).
They are solved with an extensively modified 
1D spherical version of the Eulerian code ZEUS 
(Stone \& Norman 1992).
The pressure-free dark matter and the baryonic gas
are computed as two separate fluids which 
interact gravitationally.
The size of spatial zones increases 
logarithmically. 
We use ``outflow'' boundary conditions at 
the outermost spherical zone 
where the Hubble flow receeds supersonically 
out of the grid, and the flow velocity 
vanishes at the origin.
Sink terms for radiative cooling and distributed mass 
dropout are included in the appropriate models.
For those models involving gas flow within 
optical galaxies, the gas dynamic equations 
equations have additional
source terms for stellar mass loss and heating 
by stars and Type Ia supernovae.
The full set of equations are described in 
Brighenti \& Mathews (1999a).

\subsection{External Pre-heating}

The characteristic virial temperature of an 
NFW halo can be found directly from the condition 
for hydrostatic equilibrium,
$T \sim (\mu m_p/k) G M_{vir}/r_{vir} \propto r_{vir}^2$
since by definition the mean density within the 
virial radius is always $\Delta_c \rho_c$.
The X-ray luminosity for pure bremsstrahlung 
emission, with emission coefficient 
$\Lambda \propto T^{1/2}$, is 
$L \propto \rho^2 \Lambda  r_{vir}^3
\propto \rho^2 T^2 \propto T^2$
where the characteristic gas density 
$\rho$ is assumed to scale with the dark halo density 
(e.g. Kaiser 1986; 1991).
Therefore we expect $L \propto T^2$ provided the 
following assumptions hold:
(1) the gas is in hydrostatic equilibrium bound to  
homologously identical halos,
(2) the gas density profiles are homologous 
in the variable $r/r_{vir}$ for all clusters,
(3) the ratio of gas to dark mass is identical
within all $r_{vir}$, and 
(4) X-ray line emission is ignored.
Although reasonable, these assumptions are not 
perfectly satisfied. 
For example, the gas variables 
within the accretion shock $r_s$, 
which typically occurs very close to the virial radius, 
are not expected to be strictly 
homologous in $r/r_{vir}$ and, furthermore, the 
halo concentration varies somewhat with $M_{vir}$.
Non-gravitational heating of the gas in excess of 
that received in the accretion shock can cause 
the mass fraction in gas to vary with virial mass. 
Line emission becomes important at 
$T \lta 10^7$ K, relevant for galaxy groups.
At these low temperatures 
the radiative cooling coefficient 
$\Lambda(T,Z = 0.4) 
\propto T^p$ where $p \lta 0$, which flattens 
the $L - T$ slope to at least $L \propto T^{3/2}$.
Finally, the NFW halo for a given virial mass 
and cosmology represents a computational mean 
configuration about which there is considerable 
scatter; for dark halos of the same $M_{vir}$, 
Bullock et al. (2000) find a 1$\sigma$ scatter of $\sim 50$ 
percent in the concentration parameter.

We consider four levels of pre-heating 
for each of the three virial masses. 
This is done by arbitrarily raising the temperature 
to $T_h$
everywhere in the $\Lambda$CDM flow at time 
$t_h = 0.5$ Gyrs (redshift $z = 9$) when the 
cosmic baryonic density is 
$\rho_h =  3.61 \times 10^{-28}$ gm cm$^{-3}$.
For comparison with other recent discussions 
of the pre-heating hypothesis, we also reset 
the temperature and density to be everywhere 
constant at time $t_h$.
By this time the NFW dark matter halos have 
masses about $6 \times 10^{12}$, $4 \times 10^{13}$ 
and $10^{14}$ $M_{\odot}$ for groups, poor clusters 
and clusters respectively.
Gas that had already entered the dark halos 
at time $t_h$ 
is spread uniformly with the rest of the 
gas at the reset. 
The heat supplied to the gas at these early 
times places the universal gas on an adiabat 
which can be represented with an entropy factor 
$$K_{34} \equiv 10^{-34} {kT_h \over \mu m_p \rho_h^{2/3}} 
= 2.3 \times 10^{-8} T_h = 0.31~T_{h,keV}$$
as defined by Tozzi \& Norman (2000)
which is evaluated at time $t_h$.
The corresponding entropy factor 
defined by Ponman, Cannon \& Navarro (1999) is 
$$S \equiv {T \over n_e^{2/3}} = 985 K_{34}
~~{\rm keV}~{\rm cm}^2.$$
The four levels of heating we consider -- 
$T_h = 10^4$, $5 \times 10^6$, $10^7$ and 
$3 \times 10^7$ K 
corresponding to 0.0013, 
0.65, 1.3, and 3.9 keV/particle -- are numerically 
characterized by 1-4 in our model nomenclature
and correspond to entropies 
$K_{34} = 2.55 \times 10^{-4}$, 0.127, 0.255, 
and 0.764 respectively.
The three virial masses are referred to as 
``g'', ``pc'' and ``c'' for group, poor cluster 
and cluster respectively. 
For example, model ADg1 
refers to an adiabatic model for gas 
flow in a galaxy 
group potential in which 
the temperature is set to $T_h = 10^4$ K at $t_h$.
In all model calculations, as the gas evolves 
we never allow its temperature 
to drop below $10^4$ K, as would be expected 
in a photoionized intergalactic medium (IGM). 

Previous discussions of pre-heating have not 
emphasized the possibly important 
role of Compton cooling against the cosmic background
radiation. 
After pre-heating, 
the electron gas is Compton cooled 
but the proton temperature 
$T_p$ remains unchanged  
until equipartition is established 
by Coloumb interactions.
To explore this we assumed $T_h = 10^7$ K for 
both electrons and protons at 
($t_h$, $z_h$) = (0.5 Gyrs, 9) and 
integrated equations for $T_e(t)$ and $T_p(t)$ in 
our assumed $\Lambda$CDM universe, 
allowing for Compton cooling, equipartition 
and normal expansion cooling 
$dT_{exp}/dz = 2T_{exp}/(1+z)$.
While the electrons rapidly 
lose half of their thermal 
energy relative to $T_{exp}$ 
by ($t$, $z$) = (0.64 Gyrs, 7.7), 
$T_p = 0.5 T_{exp}$ occurs at a much later time,
($t$, $z$) = (3.1 Gyrs, 2.0). 
By this latter time the proton temperature 
$T_p = 4.5 \times 10^5$ K is much less than 
halo virial temperatures of interest, 
and has negligible thermal energy as it enters the 
accretion shock.
Although our computed accretion flows are not greatly 
influenced by Compton cooling,
it should be noted that the epoch of 
(universal) 
pre-heating cannot be much earlier than 
$t_h \approx 0.5$ Gyrs or the gas would rapidly 
cool and may also leave an observable perturbation 
on the cosmic microwave background.
With this potential difficulty in mind, 
we shall assume $t_h = 0.5$ Gyrs in most of our 
calculations and consider a single temperature gas.

\section{GAS DYNAMIC MODELS FOR GROUPS AND CLUSTERS}

\subsection{Adiabatic Models}

In our ``adiabatic'' AD models the entropy only 
increases when the gas is shocked.
By far most of the entropy increase occurs when 
the gas passes through the accretion shock 
as it enters the NFW halo, 
$r_{sh} \approx r_{vir}$, and encounters 
nearly stationary gas that arrived earlier. 
The post-shock temperature is comparable to 
the halo virial temperature 
$$T_{vir} \approx \gamma{\mu m_p G M_{vir} 
\over  k r_{vir}} \approx 
2.97 \times 10^6 \left( {M_{vir} \over 10^{13}M_{\odot}}
\right)^{2/3}~~{\rm K}$$
where $\mu = 0.61$ is the mean molecular weight 
and $\gamma \approx 0.5$ from 
Eke, Navarro \& Frenk (1998).
The current virial temperature for the three 
dark halos we consider are
$T_{vir} \approx 8.1 \times 10^6$, $2.3 \times 10^7$,
and $7.2 \times 10^7$ K.
If the 
temperature of gas entering the 
dark halo $T_{pre}$ is comparable to (or significantly 
exceeds) $T_{vir}$, then the accretion shock 
weakens and gas 
flows nearly adiabatically into the halo, 
preserving its entropy.
The temperature of gas entering the accretion 
shock at the present time ($t_n = 13$ Gyrs) 
has cooled by adiabatic expansion to 
$T_{pre}(t_n) = T_h(\rho_n /\rho_h)^{2/3} = 0.01 T_h$ 
or $10^4$ K, whichever is larger.
In any case, a shock must always
propagate away from the origin (where the flow
velocity vanishes), even if the converging
flow there is subsonic.
Finally, for models with the least amount of 
IGM pre-heating (AD1), spurious shocks may develop
just after time $t_h$ as gas with sub-virial 
temperatures flows into the dark halo. 

\subsubsection{Adiabatic Models in the $L-T$ Diagram}

Bolometric X-ray luminosities and 
emission-weighted mean temperatures 
for 12 adiabatic models 
(3 $M_{vir}$ and 4 $T_h$) at time $t_n = 13$ Gyrs 
are shown with the observational data in Figure 1a. 
The gas has been heated to $T_h$ and reset to 
spatial uniformity at $t_h = 0.5$ Gyrs.
As the virial mass increases from groups to 
clusters (filled circles $\rightarrow$ squares 
$\rightarrow$ triangles), the computed results 
follow the general trend of the observations,
even for these simple adiabatic models.
For each virial mass the final luminosity is 
progressively lower as the amount of assumed 
pre-heating increases, AD1 $\rightarrow$ AD4.
This is expected since with increased pre-heating 
the entropy of the 
gas that enters the halo at early times 
may exceed 
the entropy that would have been achieved in 
the accretion shock.
As the entropy at the center of the flow 
increases, the gas density and X-ray 
emissivity are reduced, corresponding 
to lower $L$ in Figure 1a. 
With increasing virial mass, 
each cluster of four models (1 $\rightarrow$ 4)
becomes more compact in Figure 1a 
since $T_{pre} \ll T_{vir}$ is 
satisfied for the largest $M_{vir}$, even 
with the maximum amount of pre-heating. 
Further details of the AD models are listed 
in Table 1.

Although the adiabatic series of models does 
not lose energy by radiation, to compute
the luminosities plotted in Figure 1a, 
it is necessary
to assume that they do in fact radiate
with emissivity $\propto \rho^2 \Lambda(T)$. 
The cooling coefficient $\Lambda(T,Z)$ that we use 
(taken from Sutherland \& Dopita 1993) corresponds 
to abundance $Z = 0.4$ (solar meteoritic) 
for which $\Lambda(T)$ reaches a minimum 
at $T \approx 1$ keV because of increasing 
line emission at lower temperatures.
%[THE FOLLOWING IS NO LONGER TRUE!  WHY NOT?
%This explains why the series of models 
%ADc1 $\rightarrow$ ADpc1 $\rightarrow$ ADg1 does 
%not exhibit the selfsimilar, pure bremsstrahlung 
%slope $L \propto T^2$ but is instead 
%flatter at low temperatures corresponding 
%to $T_{vir}$ for groups.
%A similar flattening is also seen in 
%the uppermost curve of Figure 18 
%of Knight \& Ponman (1997). 
%Further details of these adiabatic models are 
%listed in Table 1.]

Finally, the results shown in 
Figure 1a are sensitive to 
the time $t_h$ at which the reset is made and 
are very different if the reset is ignored altogether.
For example if $t_h = 2$ Gyrs instead of 0.5 Gyrs, 
the temperatures of the final ADg models are
relatively unchanged, but the luminosities 
are very much lower than those in Figure 1a: 
$\log (L/10^{45}) = -2.48$ for ADg1 and -5.17 for ADg4. 
When the reset occurs at lower redshifts, 
more gas must be removed from the halos 
and the subsequent accretion in the $\Lambda$CDM 
universe is insufficient to re-establish the 
present-day X-ray luminosity.
Therefore explanations for both the amount of 
universal pre-heating and the epoch at which it 
is applied are required in external heating scenarios.
If the reset is completely ignored, values of 
$L$ are increased enormously.
For example, 
$L \gta 10^{46}$ erg s$^{-1}$ for model ADg1nr
(``nr'' signifies no reset), as shown in Table 1.
These no reset groups have X-ray luminosities 
comparable with the richest clusters observed.
The reason for this is that dense, low entropy 
gas flowed into the group potential 
before time $t_h$ and, if not removed, must remain 
there in the absence of reset.
The very dense central gas $n_e \sim 100$ cm$^{-3}$ 
that entered the halo at very high redshifts 
accounts for the enormous $L$ of AD groups without reset.
In more realistic models this gas would have radiated 
its energy away and formed into stars.

\subsubsection{Adiabatic Models in the $S-T$ Diagram}

Ponman, Cannon \& Navarro (1999) 
and Lloyd-Davies, Ponman \& Cannon (2000) have emphasized 
a related non-selfsimilar 
deviation of galaxy group observations 
in the entropy-temperature plane. 
They compare the emission-weighted 
mean temperature $T$ to the entropy factor
$S_{0.1} = T/n_e^{2/3}$ evaluated at $0.1 r_{vir}$, 
approximately the outer extent of current observations. 
The same idealized, self-similar group-cluster models that 
correspond to $L \propto T^2$ would 
appear as $S_{0.1} \propto T$ in Figure 2a.
However, it is seen that 
the entropy observed in groups generally exceeds the 
$S \propto T$ variation when extrapolated from 
observations of rich clusters. 
Nevertheless, the deviation from similarity 
is less robust in the $S - T$ plane 
than in the $L - T$ diagram since 
$S_{0.1} \equiv S(r = 0.1r_{vir})$ refers to a single 
point in the observed or computed entropy 
distribution and, for the observations, 
also requires an estimate of the virial radius.
The observational errors are substantial and 
can be found in the papers cited above. 

Values for $S_{0.1}$ from the computed 
adiabatic models are plotted 
as open circles in Figure 2a. 
Entropies $S_{0.1}$ 
for models AD4 with the highest level of pre-heating, 
shown with the largest circles, 
depend weakly on $M_{vir}$ 
since the entropy acquired from pre-heating 
dominates that from shocks.
As the pre-heating at time $t_h$ is reduced, 
$S_{0.1}$ decreases with 
$M_{vir}$, approaching the expected $S \propto T$.
However, $S_{0.1}$ for the least-heated models, 
ADpc1 and ADc1, 
exceeds the entropy for ADpc and ADc models that 
received more pre-heating.
The source of this discrepancy can be 
traced to post-shock irregularities in the entropy 
profile $S(r)$ 
created by weak shocks produced during the transient 
adjustment following the reset time $t_h$. 
Immediately after 
reset, cosmically inflowing gas outside the 
virial radius encounters stationary gas within.
This gives rise to both outward and inward propagating 
shocks. 
The latter shock confronts a third outward moving shock
produced as gas of 
uniform (post-reset) 
density falls toward the origin, reflecting as a shock. 
The local increase in entropy depends on the local strength 
of each shock that passes through.
These artifacts of the reset assumption imprint long-lived 
entropy irregularities in the gas. 
The effect of these transient shocks is 
much reduced in models with more pre-heating 
since the sound speed is larger and 
shocks are greatly reduced in strength. 

The resulting fluctuations in $S(r)$ are evident 
in Figure 3 where we plot entropy profiles  
for groups and clusters (ADg \& ADc) at the present time. 
For each $S(r)$ profile there are several 
regions of interest: the relatively flat 
cosmic inflow region at the far right, 
the sudden vertical entropy increase at the 
accretion shock, a region of declining 
post-shock entropy, and an inner region 
where the entropy may decline further or 
be approximately constant. 
Note also the decrease in the pre-shock entropy 
as gas approaches the shock 
when there is no pre-heating, AD1. 
This is due to our lower limit on the
inflowing gas temperature, $T = 10^4$ K. 
Since a low-temperature, 
adiabatically expanding universe would 
cool below this value, our lower limit on 
the temperature results in isothermal intergalactic flow
in which the entropy factor increases as 
$\rho^{-2/3}$. 
For this isothermal pre-shock flow, 
the entropy decreases with rising gas density 
as the flow converges toward the shock 
in the AD1 flows.
For a given dark halo, 
when the accretion shock is strong, 
the entropy just behind the shock is the same 
for all levels of pre-heating.
The post-shock flow behaves approximately 
as $S \propto r$, indicating that the accretion 
shock was weaker in the past when the 
dark halo was less massive.
In model ADg1 with no pre-heating 
this decline in the entropy continues to small radii;
the post-shock entropy was very low 
in the early universe.

When the pre-heating is large, 
for some time after $t_h$ gas enters the dark halos 
without shocking, conserving $S$; this 
constant entropy gas 
is currently found closer to the center 
in the ADg2 - ADg4 solutions in Figure 3.
The entire entropy profile for model ADg4 is 
almost constant.
However, converging subsonic flow must always 
produce an outward propagating shock since the flow 
velocity is assumed to vanish at the origin. 
The strength of this shock, which increases with virial 
mass $M_{vir}$, can produce central regions 
with $dS/dr < 0$, as in the ADc1 - ACc4 
models in Figure 3.
Since these small inner regions are convectively 
unstable, the entropy would also become uniform 
in a more realistic three dimensional calculation.

The choice of $S_{0.1}$ and emission averaged 
temperature $T$ in the $S - T$ diagram 
are determined by the realities of 
observational limitations, but they are probably not 
the best coordinates to test deviations from similarity. 
These inner flow regions near $0.1r_{vir}$ are 
also subject to computational difficulties 
from ancient shocks as 
apparent in the AD1 and AD1nr entropy profiles
in Figure 3.
Even if the cosmic gas were perfectly adiabatic 
between shocks, 
we expect that the violent events that accompany the 
earliest mergers would disturb the entropy of the 
central regions near $0.1 r_{vir}$ which in 
our models 
passed through the accretion shock at $\sim 1$ Gyr
(redshift $z \sim 5$).
The emission-weighted temperatures also heavily 
favor the same dense inner region sensitive 
to these early perturbations.
Aside from the entropy fluctuations 
apparent in the inner flow $r \lta 0.1r_{vir}$ 
in Figure 3,
our calculations are quite accurate elsewhere.
This is evident in Figure 4 where we plot 
entropies and temperatures that are mass-averaged 
values within the accretion shock. 
In these $S - T$ coordinates 
the unheated AD1 models lie exactly along the 
similarity variation $S \propto T$. 
Small deviations from this relation occur with increased 
pre-heating and smaller virial masses, 
just as one would expect.

When the reset assumption is not made, 
as in the ``nr'' solutions listed in Table 1 
and ADg1nr and ADc1nr plotted in 
Figure 3, the entropy in the inner post-shock flow is 
very much lower.
In these flows gas at $10^4$ K 
that entered the halo at high redshifts, when 
the cosmic flow was dense, had very low entropy.
The high density of this gas is maintained 
during its evolution.
In addition the emission-averaged temperatures of
the ``nr'' solutions are much too low, biased 
by the central regions.
This is still true even if all the gas is heated 
(without reset)  
to $3 \times 10^7$ K at time $t_h = 0.5$ Gyr, 
as shown in the ADg4nr, ADpc4nr and ADc4nr results 
in Table 1. 
Some irregularities in the entropy profile in 
Figure 3 for the ADc1nr flow arise from 
weak shocks produced at very early times 
$t_{tr}$ when the 
dark halo mass distribution changed to an 
NFW profile. 

Clearly, adiabatic models without reset are unacceptable 
since their X-ray luminosities are excessive and 
they have not cooled or formed into stars and galaxies.
The high luminosities of no reset flows are related to the 
strong high redshift emission from groups that 
would violate the soft X-ray background, as discussed by 
Pen (1999) and Wu, Fabian \& Nulsen (1999b).
But adiabatic solutions with the reset condition are
also problematic because the time $t_h$ of universal 
heating must be carefully orchestrated for 
agreement with observed X-ray luminosities of galaxy groups.
Nevertheless, AD flows provide a useful reference 
for more detailed externally heated flows. 
Moreover, the entropy profiles $S(r)$ for the AD solutions 
in Figure 3 share many 
of the same features with more realistic 
externally pre-heated models discussed below.

\subsection{Models with Radiative Cooling}

This series of gas dynamic models, 
described with the prefix CO, is similar to 
the adiabatic models except we now include 
radiative emission according to the
Sutherland-Dopita cooling coefficient
$\Lambda(T,Z=0.4)$.
When the gas cools at 
the center of the flow, we allow it to accumulate there, 
producing a point mass gravitational potential similar to 
a giant black hole. 
At time $t_h = 0.5$ Gyr we reset the gas temperature and 
density to be spatially uniform and apply the 
pre-heating.
These assumptions are essentially 
identical to those made by Tozzi \& Norman (2000).
The entropy decrease due to radiative losses 
has little effect on the temperature profile 
since for hydrostatic support 
the gas temperature 
must always be close to $T_{vir}$, 
i.e., $T \propto M(r)/r$ and 
$M(r)$ is approximately proportional to $r$.
However, close to the central 
concentration of cooled gas, 
$M(r)$ is nearly constant and the 
temperature increases toward the origin 
as $T \propto 1/r$. 
The locally higher temperatures in this 
dense central gas can influence the global  
emission-weighted temperature 
of the group or cluster.

In some models with radiative cooling hot gas 
containing an appreciable amount of specific enthalpy 
and kinetic energy flows into the central numerical zone. 
Since this gas must ultimately radiate this energy  
producing X-rays,
the central region contributes an additional 
X-ray luminosity given by 
$$L_{core} = 4 \pi r_1^2 \rho_1 u_1
\left( {u_1^2 \over 2} + {5 k T_1 \over 2 \mu m_p} \right)
~~~~{\rm erg~s}^{-1}$$
where the subscript 1 refers to quantities 
evaluated at the radius $r_1 = 150$ parsecs of the 
innermost numerical grid.
This estimated correction to the total luminosity 
can be quite large, often exceeding $L$ from the 
rest of the cooling flow.
Since X-ray observations of cores of cooling flows 
do not reveal strong, high-temperature thermal point 
sources, the assumption that the gas cools only at the 
center of the flow is unrealistic. 

\subsubsection{Cooling Models in the $L - T$ Diagram}

$L$ and the emission-weighted $T$ for the 
CO (radiative cooling) models are illustrated in Figure 1b.
The result for cluster mass halos 
(COc = filled triangles) 
are similar to the adiabatic models with reset, 
as a result of the inefficiency of radiative cooling 
in these hot, low-density halos.
In gas flows with the largest 
pre-heating -- COg3, COg4, COpc3, COpc4 and COc4 -- 
none of the 
baryonic gas cools in $t_h < t < t_n$, as shown 
in Table 2. 
These strongly pre-heated 
models fit the observations in Figure 1b 
quite well even for low luminosity groups. 
However, the remaining 
cooling flow models with less pre-heating 
are currently accumulating 
uncooled gas in the origin.
As shown in Table 2, 
for these flows the concentrated X-ray emission from the 
central zone $L_{core}$ is 3 - 13 times 
that from the rest of the flow within the virial radius. 
In addition, the colossal masses in the 
central baryonic singularity, $\gta 6 \times 10^{11}$
$M_{\odot}$, are similar to those of massive galaxies.
Clearly, none of these weakly pre-heated models are physically 
acceptable.

Another distinctive feature of the 
CO series of calculations is the appearance of 
galactic drips 
at early times, $\sim 1$ Gyr for clusters and 
poor clusters, and slightly later 
$\sim 5$ Gyrs for groups.
Galactic drips are dense, narrow 
cooling waves that begin in the 
outer halo and proceed inward, crossing the
cooling flow in $\lta 1$ Gyr. 
The drips discussed by Mathews (1997) actually 
lose mass en route by local cooling.
But drips in the 
CO models continue to the origin 
before cooling, so they are 
more massive and move faster 
than the drips discussed by Mathews (1997).
The cooling evolution of spherically symmetric
drips depends somewhat on the numerical resolution.
Spherical drip waves are an artifact of spherical symmetry 
in which the inherent three-dimensional character of 
Rayleigh-Taylor instabilities is suppressed. 
Although drip waves have a physically plausible 
origin and may well occur in cooling flows, 
the wave fronts are unlikely to 
be globally spherical and 
their multi-dimensional evolution is 
currently uncertain. 
Drips occur only in CO models 
with little or no pre-heating 
(e.g. COg1, COc2, COpg1 and COc1) 
and only one or at most two drip waves 
occur in each of these models.
As drips arrive at the origin,
the mass of cooled gas there
($M_{cool}$ in Table 2) suddenly increases.
After a drip wave has passed through gas 
at the center of the flow, its entropy increases 
because the drip waves 
become supersonic in $r \lta 0.1 r_{vir}$. 
The post-drip gas density is also lower 
since the drip wave transfers mass to the center.
These factors contribute somewhat to 
the low $L$ for the COg1 model in Figure 1b.
Both density and temperature gradients in the 
inner flow ($r \lta 0.1 r_{vir}$) 
are considerably flattened 
following the passage of drip waves. 
Since drips typically begin at $r \lta 0.3 r_{vir}$,
and move inward; they have little influence 
on the outer $\sim 70$ percent of the 
cooling flows.

\subsubsection{Entropy of Cooling Models 
\& No Reset Models}

Entropy factors $S = T/n_e^{2/3}$ (evaluated 
at $r = r_{vir}/10$) for the CO models are plotted 
against emission-averaged temperatures 
for groups and clusters in Figure 2b. 
As the amount of pre-heating is reduced, 
the results approach the expected $S \propto T$ 
variation, 
but the entropy in these weakly pre-heated flows 
is increased by the passage of drip waves 
and shocks produced by transient flows 
following reset.
Because of these complications, the entropy $S_{0.1}$
computed at small radii is uncertain and the 
agreement with observations in Figure 2b is poor.
The globally mass-averaged entropy and temperature 
satisfy the similarity condition 
$S \propto T$ very well (Figure 4b), 
but cannot be easily compared with 
currently available observations.

In Figure 3 we show the computed entropy
profiles $S(r)$ at $t_n = 13$ Gyrs
for group and cluster CO models.
In general these entropy profiles are in good 
agreement with the approximate profiles 
computed by Tozzi \& Norman (2000), 
but models that experienced drips
(e.g. COg1) have higher central entropies 
and gas temperatures in $r \lta 0.1 r_{vir}$ 
due to the influence of drip waves. 
For the most strongly pre-heated models 
the entropy is high everywhere. 
In less strongly pre-heated flows 
the entropy is seen to decrease slightly 
as the gas approaches the accretion shock; 
this occurs 
because the pre-shock flow is isothermal 
at $T_{pre} = 10^4$ K,
as explained earlier. 
The accretion shock (at all times) 
occurs very close to the virial 
radius: $\log r_{vir}({\rm pc}) = 2.94$ and $3.42$ for 
groups and clusters respectively.
The entropy just behind the strong accretion shock 
increases with virial mass.
As with the AD solutions,
the positive entropy gradient in the
immediate post-shock flow,
$S \propto (r/r_{vir})^p$ with $p \approx 1$, 
is due to the
increasing strength of the accretion shock
with cosmic time (Tozzi \& Norman 2000).
In the inner flow region, $r \lta 0.1r_{vir}$
the entropy variation $S(r)$ depends on 
several competing processes.
The central entropy is increased by 
drip-induced shocks (COg1) or 
(in COg3 and COg4) by 
outward-propagating shocks 
generated by the initial nearly adiabatic 
flow of uniform post-reset 
gas into the origin.
In this latter case $dS/dr < 0$ so the 
(small mass of)
gas very close to the center of the flow 
would be convectively unstable.
Radiation losses from the dense central regions 
lower the central entropy of CO models 
below the corresponding AD model, although 
the effect of this can be reversed by drips 
(COg1, COc1, and COc2).

We have also computed representative CO flows 
with no reset, COnr models, in which the gas 
acquired by the dark halo before $t_h$ is not removed.
In these exploratory calculations 
the temperature of CO4nr models is increased by  
$3 \times 10^7$ K at $t_h$ but CO1nr 
models receive no additional heating.
The unsatisfactory results of these models 
are listed in Table 2. 
All of these ``nr'' flows suffer from early intense 
radiative (over)cooling, producing baryon mass singularities 
about ten times larger than those of the AD models.
As a result, the X-ray luminosity is 
currently totally dominated 
by emission from the core, $L_{core} \gg L$.
The unrealistic nature of these COnr models 
without reset strongly
argues for star formation and supernova heating
that drives
gas out of small dark halos, locally raising its 
entropy.

In summary, 
our solutions for the CO models are in good 
agreement with those of 
Tozzi \& Norman (2000), particularly in the post-shock 
entropy profiles shown in Figure 3.
But our gas dynamical calculations have illustrated 
some additional features:
the presence of drip waves, the concentrated 
X-ray emission from the baryonic singularity, 
a density enhancement 
as the gas approaches the accretion shock 
(Tozzi, Scharf \& Norman 2000) 
and shock heating at the origin. 
Our CO results depend 
critically on the  
assumption used by many authors 
(e.g. Kaiser 1991; Cavaliere et al. 1997; 
Balogh, Babul \& Patton 1999; 
Wu, Fabian, \& Nulsen 1999a;
Tozzi \& Norman 2000)
that the baryonic density and temperature are 
nearly uniform at the moment of pre-heating 
$t_h$ before gas flows 
into the dark halo potentials. 
In fact, however, baryonic gas is already 
concentrated in the dark halos at time $t_h$ 
and the most natural means of removing it is 
with supernova-driven winds. 

\subsection{Mass Dropout Models}

The DO calculations 
are identical to the CO models 
except the cooling flows are assumed to be inhomogeneous, 
containing entropy (or magnetic) irregularities that allow 
the gas to cool (``dropout'') at large distances from the origin. 
These DO models are an improvement over the 
CO models in which baryonic mass singularities form 
at the origin having masses 
that are much greater than the masses 
of black holes observed in luminous elliptical galaxies. 
The CO models also produce extremely bright 
central X-ray sources that are not observed. 
Although the details of mass dropout are poorly understood, 
spatially distributed radiative cooling 
is required in elliptical galaxy cooling flows.
For example, 
the masses of central black holes in elliptical galaxies 
are about 10 times smaller than the total amount of 
diffuse cooling flow gas that has cooled over a Hubble time.
In addition, 
the X-ray isophotes of cooling flow gas 
in rotating galaxies does not exhibit 
rotational flattening; 
this can be understood if angular momentum is being 
lost by distributed radiative cooling
(Brighenti \& Mathews 2000a,b) 

The local cooling dropout rate in an inhomogeneous 
cooling flow depends on the 
amplitude distribution of the entropy or magnetic inhomogeneities, 
which cannot be directly observed or derived {\it ab initio}. 
For a simple heuristic representation of mass dropout 
it is customary to introduce 
a sink term in the equation of continuity of the form
$-q(r) \rho /t_{do} \propto \rho^2,$ 
where $t_{do} = 5 m_p k T / 2 \mu \rho \Lambda$
is the time for gas to cool locally by radiative
losses at constant pressure. 
With this term the elliptical galaxy 
X-ray surface brightness 
distributions more nearly resemble those observed and 
the dynamical mass to light ratios are not 
greatly disturbed (e.g. Sarazin \& Ashe 1989;
Brighenti \& Mathews 2000b). 
Successful models require 
that the dimensionless dropout parameter $q$ is close 
to unity.
We assume $q = 1$ in the DO models computed here 
and it applies only to hot gas $T > 10^5$ K.
The influence of 
galactic drips is lessened in the DO models, but 
not entirely eliminated; their amplitude and velocity 
are reduced as gas cools in the wave 
and is locally deposited, similar to 
the drips described by Mathews (1997).
Since we view the DO models as a variant of the CO models, 
we continue to reset the gas temperature 
and density at time $t_h$. 
Flow irregularities introduced by the reset 
tend to be lessened by mass dropout in the DO models.
Finally, we note that when $q = 1$, 
about half of the bolometric 
X-ray luminosity is produced by the cooling regions
(Brighenti \& Mathews 1998). 

An important beneficial effect of the distributed dropout 
models is 
that they do not produce the unrealistic central baryonic 
mass concentrations that occur in the CO models.
Instead, the cooled baryons are assumed to remain 
at approximately the same radius at which the cooling 
dropout occurs.
This would be expected if the cooled gas forms into 
stars that spend most of their time orbiting 
near the radius where they formed.
As gas cools according to the $q = 1$ sink term over 
many Gyrs, an extended region of stars 
forms from the cooling dropout (Nulsen \& Fabian 1995, 1997). 
The density structure of the dropout stellar population 
has a remarkable resemblance to a de Vaucouleurs profile, 
particularly at early times, $t \sim 2$ Gyrs.
For the DO models discussed here the dropout term is 
applied to the baryonic gas within the current virial 
radius; at earlier times only a very small amount 
of dropout occurs in the low-density pre-shock cosmological flow.

\subsubsection{Dropout Models in the $L - T$ Diagram}

The results of the spatially distributed dropout 
(DO) models in the ($L - T$)-plane 
are shown in Figure 1c. 
The DO models exhibit many of the same trends as the CO 
models in Figure 1b but the cooled baryons are less 
concentrated so the X-ray surface brightness distribution 
is also less centrally concentrated. 
For each of the three virial masses we consider, 
the total mass of cooled gas in the DO models $M_{cold}$ 
within the current virial radius 
is insensitive to the amount of pre-heating,
as shown in Table 3. 
The filled triangles in Figure 1c, corresponding to 
models with cluster mass 
halos, DOc, are in excellent agreement with observations and  
are largely unaffected by the various levels of 
pre-heating. 
The poor cluster solutions lie above the Arnaud-Evrard 
observations, but recall that their sample favored 
clusters with weak cooling flows 
which are hotter and therefore appear 
systematically underluminous in the $L - T$ plot. 
Also Arnaud \& Evrard did not 
extrapolate the observed X-ray luminosities 
to the virial radius.
For both groups and poor clusters the luminosity 
decreases as the amount of pre-heating increases. 
The emission-weighted temperatures also decrease slightly 
as DO1 $\rightarrow$ DO4 
since the more strongly pre-heated models produce less 
total mass dropout and the lowered gravity of this 
dropout mass requires lower gas temperatures for 
hydrostatic support. 

Overall, the results in Figure 1c for 
group, poor cluster and rich cluster DO models are in 
good agreement 
with the observations, particularly if 
the pre-heating is strong. 
Models DOg1 - DOg3, with little or no pre-heating, clearly 
lie above the group 
observations of Helsdon \& Ponman (2000a).
The most consistent models, when compared 
to the observations, are DOg4, DOpc4 and DOc4. 
While this clearly supports 
the argument for cosmic pre-heating that 
has been widely discussed, 
we note that the amount of pre-heating 
required cannot be explained solely by supernova 
heating associated with normal star formation 
(e.g. Loewenstein 2000).
The timing is also wrong since most galactic stars 
are thought to have formed  
at redshifts much less than $z_h = 9$, corresponding 
to the epoch of pre-heating $t_h = 0.5$ Gyr.

Since the DOg4 model agrees so well with typical 
Heldson-Ponman groups in Figure 1c, 
it is interesting to estimate the amount of pre-heating 
that would be required to raise the entropy to the 
DOg4 (preshock) level at various redshifts.
The amount of heating required to reach the 
same entropy of the DOg4 model is 
3.9, 0.92, 0.59, 0.33 and 0.15 keV/particle if the 
heating occurred at redshift $z = 9$, 4, 3, 2 and 1 
respectively.
If about 0.1 of all baryons formed into stars 
($\Omega_*/\Omega_b = 0.09$ 
Fukugita, Hogan \& Peebles 1998) 
with a Salpeter IMF (as discussed above), 
this would generate only about 0.19 keV/particle.
The required level of cosmic entropy in model 
DOg4 (Figure 1c) would be (just) consistent with 
supernova heating and normal star formation 
if most stars formed at redshifts $z \lta 1$.
%although the cosmic gas is certainly no longer 
%uniform at these low redshifts. 
Such recent supernova 
heating is clearly inconsistent with 
our knowledge that most stars in 
group-dominant early type galaxies 
were formed well before redshift $z = 1$. 
Therefore, if the pre-heating is cosmic and universal it 
must have a non-stellar (AGN?) 
origin or be produced by (Pop III?)
stars that have no surviving IMF counterparts today.
In addition, the pre-heating in our models 
occurred at $t_h = 0.5$ Gyr 
before very much gas entered the dark potential 
(e.g. Wu, Fabian \& Nulsen 1999b); 
if the heating by stars occurred at later 
times, much luminous gas would have entered 
the dark potentials, in possible conflict with the 
observed X-ray background.

\subsubsection{Entropy, Cooled Gas \& No Reset 
in DO Models}

The behavior of 
DO models at time $t_n = 13$ Gyrs in the $S-T$ plot is 
illustrated in Figure 2c. 
For each virial mass the entropy decreases 
with decreasing amounts of pre-heating. 
However, the DOg models that fit best in the $S-T$ plot, 
DOg1 and DOg2, agree less well with observations 
in the $L-T$ plot, Figure 1c; 
the best compromise model may be DOg3.
However, we note again that the $S-T$ plot 
(with $S$ and $T$ evaluated at $r= 0.1r_{vir}$)
is a less
accurate indicator of the amount of pre-heating than 
the more globally representative $S-T$ plot
in Figure 4c or the $L-T$ plot (Figure 1c).

Figure 3 shows the final entropy profiles 
in the group and cluster DO models.
Many of the same features shown for the CO models 
appear again, but there are some significant 
differences, particularly in the inner regions 
$r \lta 0.1r_{vir}$.
Most of these differences can be understood in terms 
of the gas velocity.
In the CO solutions, the gas accelerates within 
$0.1 r_{vir}$  
as it approaches the central mass concentration, 
so less energy is radiated in this region 
and the entropy profiles become flat.
By contrast, in the DO solutions 
the central gas velocity is slowed by mass dropout 
and a much larger fraction of its entropy is radiated away.
Dropout causes subsonic flows to move even slower
(Sarazin \& Ashe 1989; 
Brighenti \& Mathews 2000b).
This explains the continued decrease in $S(r)$ 
in many of the DO models as gas flows from 
$\sim 0.1r_{vir}$ to the origin.  
For the DOg models this region would be dominated by 
the central elliptical galaxy.
Note that the group flow without 
pre-heating, DOg1, contains a low-entropy 
drip wave at $\log r = 2$ which is moving slowly 
inward, but it has no strong influence on the solution 
elsewhere. 
Radiation losses are less important in 
dropout flows with the largest pre-heating, 
DOg3, DOg4 and DOc4, and the entropy remains 
more nearly constant throughout the inner flow.

Cooled mass profiles for representative DO group and 
cluster models are shown in Figure 5 and 
the total dropout mass within the virial radius
$M_{cold}$ is listed in Table 3.
It is remarkable that 
the dropout mass distributions for the groups 
(DOg1 \& DOg4) resemble the de Vaucouleurs profile 
of NGC 2563 shown with a dotted line in Figure 5. 
The total dropout mass $M_{cold}$ 
is also similar to total stellar masses 
of galaxy groups. 
The similarity with a de Vaucouleurs profile is 
coincidental, since 
hierarchical merging can also do this and is 
more plausible physically.
Moreover, 
in a more realistic model the mass dropout profiles 
would be lowered by internal supernova 
heating (feedback) not included in the DO models. 
The dropout profiles in Figure 5 
for groups are almost identical
for all models within $\log r_{kpc} < 0.5$ since this gas 
cooled before $t_h$ when all the models 
for given $r_{vir}$ were identical.
In the more important interval 
$0.5 \lta \log r_{kpc} \lta 2.1$ all dropout profiles 
exceed that of the profile just before pre-heating 
at $t_h$ (dashed-dot line) which 
for DOg($t_h$) terminates at 
$\log r_{kpc} = 2.1$, the location of the accretion shock
at $t_h$.
In this range we see that more dropout mass is deposited
in models with less pre-heating (DOg1) or with 
no uniform reset at $t_h$ (see below). 
The dropout profiles for cluster flows behave in 
a qualitatively similar manner, but are 30 - 100 times
denser than NGC 2563 in $\log r_{kpc} \lta 1.2$.
The dropout density in this region can be reduced 
by $\sim 10$ (without much influencing $L$ and $T$) 
if the initial tophat perturbation 
is more extended, less dense and more massive.
Nevertheless,
the high density mass dropout cores in DO cluster models 
indicate that our mass dropout assumptions 
are inappropriate, at least on cluster scales.
Finally, 
each of the mass dropout profiles has a small density 
peak at $\log r_{kpc} \sim 0.6$ kpc; this feature 
is an artifact of mass dropout that accompanied 
the first compression and heating of baryonic 
gas in the tophat perturbation. 

DO models are also sensitive to the epoch of
pre-heating
For example, when $t_h$ is increased to 2 Gyrs, 
the final luminosities are reduced 
by factors of 6, 30, 75 and 140 for 
models DOg1 $\rightarrow$ DOg4.
We have also computed ``no reset'' dropout models 
in which the temperature is increased 
at time $t_h = 0.5$ Gyr without altering the gas density 
profiles.
In models DO1nr, DO2nr, DO3nr and DO4nr the temperature 
was increased throughout the flow at by 
0, 0.5 $\times 10^7$, 1.0 $\times 10^7$ and 3.0 
$\times 10^7$ K respectively.
Global parameters for DOnr models for groups and clusters 
are listed in Table 3.
The locus of clusters in the ($L - T$)-plot is 
not strongly influenced by 
ignoring the reset, but the final 
group emission-weighted 
temperatures are increased by about 40 percent
(Table 3).
Emission-weighted temperatures reflect the temperature 
of gas in the high density cores.
If we had used mass-weighted temperatures, the 
temperatures for DOgnr models would have been 
nearly equal to DOg temperatures.
In general 
pre-heating is less effective in the no reset models 
since the heating is radiated away by the high gas 
densities after time $t_h$. 
As a result, the entropy profiles $S(r)$ for these 
models vary as $S \sim r$ in $r \lta 0.1r_{vir}$.

\subsection{Models with Star Formation and Central Galaxy}

In this GA series of gas-dynamical models we abandon the 
hypothesis of universal pre-heating and assume 
that all heating results 
from normal star formation inside the group or cluster.
This is similar to the 
approach we have used to reproduce the X-ray emission 
profile observed in 
the giant elliptical NGC 4472 (Brighenti \& Mathews 1999a).
The dark halos evolve just as in the 
previous models, but the intergalactic baryonic gas 
is assumed to remain at $T = 10^4$ K until it arrives at 
the accretion shock.
We assume that stars form 
at some early time, $t_* \sim 2$ Gyrs
(redshift $z_* = 3$), after enough
baryons have entered the dark potential. 
Before time $t_*$ we use the dropout model
to approximate the distribution of cooled baryonic mass. 
At time $t_*$, however, the baryons 
are rearranged into a density profile 
similar to that of a luminous elliptical galaxy. 
We use this same galaxy core at the centers 
of poor cluster and cluster calculations with 
an additional extended stellar component approximated 
with a King distribution; 
the gravity produced by these extended 
cluster stars has almost no influence 
on the gas dynamics. 
Also at time $t_*$ we release the 
appropriate SNII energy 
within the shock radius $r_{sh}$
in proportion to the local gas density. 
Shortly thereafter, a SNII-driven starburst wind 
occurs in galaxy group halos, 
rapidly expelling most of the hot gas 
from the vicinity of the central group.
A strong shock moves
upstream against the converging cosmic gas.
The gas that participated in the 
starburst wind reverses velocity and later shocks 
back onto the group for the
second time, with a further entropy increase.
For the deeper dark matter halos in poor clusters 
and clusters, the starburst cannot expel much gas 
beyond the current virial radius. 
Within the half-light radius of the central 
galaxy most of the hot gas is provided
by stellar mass loss.
For the GA series of models we include 
additional terms in the gas dynamical 
equations to allow for stellar mass loss and 
heating by stars and Type Ia supernovae; 
these modified equations are 
described in detail in Brighenti \& Mathews (1999a).

Our GA models are based on the mass distribution in 
galaxy group NGC 2563.
The central elliptical 
galaxy has luminosity $L_B = 7.44 \times 10^{10}$ 
$L_{B,\odot}$ (distance = 78 Mpc) and total 
stellar mass $M_{*t} =  7.57 \times 10^{11}$ $M_{\odot}$.
In addition, Zabludoff \& Mulchaey (1998) have identified 
about 45 galaxies in the group surrounding NGC 2563. 
Using the approximate morphological types of 
Zabludoff \& Mulchaey and 
$M/L_B$ ratios from Fukugita, Hogan \& Peebles (1998),
we estimate that the total stellar mass in 
the NGC 2563 group is at least 1.5 times that of 
the central elliptical galaxy. 
In the models discussed here we assume that 
the stellar mass of the entire group is twice 
that of the central galaxy NGC 2563. 
The contribution of outlying group members 
to the total stellar mass of groups is highly variable 
in galaxy groups and is often considerably larger than that for 
NGC 2563.  

Immediately
after the time of star formation $t_*$ 
the ratio of stellar to total baryons
within the accretion shock is
$M_*/M_b = 0.52$.
This ratio is based on the estimated stellar 
mass of the NGC 2563 group, 
but we use this same ratio in poor clusters and cluster 
flows. 
We note that the fraction of stellar baryons exceeds
the cosmic average $\Omega_*/\Omega_b = 0.09$ 
at zero redshift 
(Fukugita, Hogan \& Peebles 1998).
However, additional gas accretes into the virial radius 
after time $t_*$ and by the present time 
$M_*/M_b$ decreases to 0.27 in the GAg2 model and 
to 0.15 in models GApc2 and GAc2. 
The final ratio of stellar to gas mass inside $r_{vir}$
is $M_*/M_{gas} = 0.60$ for GAg2 and 0.23 for GApc2 and 
GAc2.
In Figure 6 we show the spatial variation of the 
relative mass in baryons and gas for models 
GAg1 an GAg4. 
The relative mass of baryons $f_b = M_b/M_{tot}$ also 
includes the dropout mass of cooled gas.
Note that the gas fraction $f_g = M_{gas}/M_{tot}$ 
is lower than the cosmic value even at the virial 
radius and that its value is sensitive to the total 
amount of SNII energy released. 

Except for the $\Lambda$CDM cosmology assumed here, 
the gas dynamical calculations, stellar mass loss rates 
and Type Ia supernova rates that we use are identical 
to those described in Brighenti \& Mathews (1999a).
We assume distributed mass dropout with parameter
$q = 1$.
The energy provided by Type II supernovae 
is estimated by assuming  
$E_{SN} = 10^{51}$ ergs is released 
from stars with initial masses $> 8 M_{\odot}$ 
and that $\eta_{II}$ Type II supernovae are produced per 
solar mass of stars formed. 
For a typical Salpeter IMF 
(slope $x = 1.35$, $m_{low} = 0.08$, $m_{high} = 100$) 
we find $\eta_{II} = \eta_{std} \equiv 6.81 \times 10^{-3}$.
The SNII energy delivered to the hot 
gas in a galaxy group or cluster 
of total stellar mass $M_{*t}$ 
is $M_{*t} \eta_{II} \epsilon_{SN} E_{SN}$ 
where $\epsilon_{SN}$ is the efficiency that 
the supernova energy is delivered to the thermal 
energy of the hot gas.
In cosmological simulations the efficiency 
of Type II supernova feedback is often chosen to be 
$\epsilon_{SN} \sim 0.1 - 0.2$, to allow for 
radiative losses.
Our calculation is somewhat different since we 
explicitly allow for radiative losses in the 
thermal energy equation, although we 
also assume that the supernova blast waves interact 
directly with the hot gas, not with cold, dense clouds 
as might be expected in star-forming regions.
We therefore consider several values of the 
composite parameter 
$$\eta^* \equiv \left( {\eta_{II} \over \eta_{std}} \right) 
\left( {E_{SN} \over 10^{51} } \right)
\epsilon_{SN}$$ 
which is expected to be of order unity.
For the GA models 
we consider four values: $\eta^* = 0.5$, 1, 2 and 4,
respectively designated as GA1, GA2, GA3 and GA4. 
Since $\Omega_*/\Omega_b$ is the same in all GA models,
The gas temperature following the deposition of SNII
energy is increased by 
$\Delta T = \eta^* 1.91 \times 10^7$ K.

\subsubsection{Galaxy Formation Models in the $L - T$ Diagram}

The GAg models 
for galaxy groups are most satisfactory 
since the stellar mass distribution and 
formation epoch $t_*$ are designed with groups in mind.
It is particularly gratifying that all GAg models  
fall very close to the observed position of
NGC 2563 in the $L - T$ plot, 
shown with a large open square in Figure 1d. 
Since we used NGC 2563 to estimate the stellar 
mass distribution for our group models, 
the agreement in Figure 1d 
may confirm the approximations we have made 
in the early evolution of groups.  
The GAg2 and GAg4 luminosities and temperatures 
are within about 10 and 25 percent of the values observed 
in NGC 2563 (see Table 4 for further details). 
All GAg models also lie near the upper envelope of the 
Helsdon-Ponman group data, where strong cooling flows 
are dominated by a massive central elliptical galaxy 
as we have assumed.
Generally, as $\eta^*$ increases along the sequence 
GAg1 $\rightarrow$ GAg4 the X-ray luminosities decrease. 
However, the non-monotonic behavior of model GAg3, 
$L({\rm GAg1}) > 
L({\rm GAg3}) > L({\rm GAg2})$,  
can be understood by 
competing influences on $L$: 
larger heating ($\eta^*$) decreases the hot gas density,
which lowers $L$, 
but as $\eta^*$ increases less gas cools 
and the larger surviving hot gas mass 
increases $L$.

Values of $\eta^*$ as large as 4 used in 
model GAg4 may not be unreasonable.
Mathews (1989) 
estimated the stellar mass to light ratio $M_*/L_V$ 
of luminous elliptical galaxies using 
single burst stellar populations 
with power law IMFs having many slopes and mass limits.
He showed that the Salpeter slope $x = 1.35$
is inconsistent with $M_*/L_V \sim 9$ which is typically 
observed in luminous elliptical galaxies, 
while shallower IMF slopes, $x \approx 0.7 \pm 0.3$, 
give satisfactory $M_*/L_V$ for all 
reasonable upper and lower mass limits.
A power law IMF characterized with 
slope $x = 0.7$, and mass limits 
$m_{low} = 0.08$ and $m_{high} = 100$, 
delivers much more SNII energy than the Salpeter 
IMF, $\eta_{II}/\eta_{std} = 3.5$, which 
may justify our largest values of $\eta^*$.

For the GAg models we have also explored varying 
the time $t_*$ when the SNII energy is released 
and the radial distribution where the SNII energy 
is deposited.
For the standard model, GAg2, we varied 
$t_{*}$ from 2 to 1.5 or 1 Gyr and found that the computed 
points in the $L - T$ plane at time $t_n = 13$ 
Gyrs were nearly unchanged. 
In addition, we repeated the GAg2 calculation
but applied all of the SNII energy only to 
(denser) gas within half 
the accretion shock radius at time $t_{*} = 2$ 
Gyrs.
The final luminosity increased slightly 
since the more central 
heated gas was denser and lost 
a larger fraction of its energy (and entropy) 
to radiation. 
We conclude that the positions of the group GAg models 
in the $L - T$ plot (Figure 1d) are not sensitive 
to the time or spatial dependence of the SNII energy 
release.

For completeness we have also computed 
GApc and GAc models for poor clusters and clusters.
These results shown 
in Figure 1d and Table 4 appear to have temperatures about 
50 percent lower than 
those of the pre-heated models or the corresponding 
observations.
While our GA assumptions are 
reasonable for group evolution, 
they are less appropriate for clusters.
For example, we assume that clusters grow from a single 
seed galaxy group in the core, 
ignoring the complex merging events by which 
groups combine to form large clusters.

\subsubsection{$S - T$ plot and Entropy Distribution for GA Models}

The group models GAg fall nicely among the observations 
in the $S - T$ plane shown in Figure 2d regardless of 
the heating parameter $\eta^*$.
This insensitivity of $S(r = 0.1r_{vir})$
to the initial SNII energy $\eta^*$ is a result of 
radiative losses in this inner region. 
Strong radiative cooling at the galactic center  
regulates the central density there 
to $n_e \sim 0.1$ cm$^{-3}$ 
and the temperature profile adjusts to maintain 
hydrostatic equilibrium in the common potential of all 
GA models. 
Therefore $S = T/n_e^{2/3}$ evolves to 
a similar value for all $\eta^*$.
The results for poor clusters and clusters are less 
satisfactory.
Some of the small perturbations of the entropy at 
$r = 0.1r_{vir}$ discussed earlier are also present in 
the GA models, but tend to be smoothed due to 
radiative losses and mass dropout. 
As before, when the entropy and temperature are 
mass-averaged within $r_{vir}$ the behavior is much 
more reasonable, as shown in Figure 4d.

Final entropy profiles $S(r)$ are shown for GAg and GAc 
models in Figure 3. 
The observed entropy profile for NGC 2563 is shown 
with the GAg models. 
The agreement is quite good although our models
are somewhat denser overall with slightly lower $S(r)$. 
As $\eta^*$ increases, the GAg models have 
progressively more 
extended regions of high entropy that project beyond 
the current virial radius. 
These entropy features are unlike any of the other 
models and, if they could be observed,
would be a clear signature of the internal heating 
scenario.
The outer GAg flows contain two outward-facing shocks.
The outer shock is the starburst blast wave that 
began at time $t_*$; the inner shock which occurs 
close to $r_{vir}$ is the accretion shock. 
Variations in SNII heating have little effect on the 
entropy profile or accretion shock radius in 
the GAc solutions because of their much greater mass.
A positive entropy gradient $dS/dr > 0$ extends 
to very small radii in both GAg and GAc solutions
in Figure 3d.
This is due to mass ejection of low entropy gas 
from galactic stars and radiative cooling in 
high density gas within the deep central stellar potential.

\section{DISCUSSION AND SUMMARY}

In the foregoing we have described the heating and 
dynamical evolution of hot gas in 
galaxy groups and clusters in a flat $\Lambda$CDM universe 
with a special emphasis on the gas entropy. 
The motivation for our interest in this problem 
can be traced to 
Kaiser (1986; 1991) who first noted that the 
X-ray luminosities of groups and clusters  
vary with gas temperature more steeply than 
that expected from bremsstrahlung 
emission in purely adiabatic flow 
with adiabatic shocks, $L \propto T^2$. 
The observed steeper variation 
$L \propto T^3$ or $T^4$ can be understood
if cosmic gas is heated by some non-gravitational 
process. 
The density and X-ray luminosity of 
hydrostatically supported gas in dark halos 
are both lower when the gas is heated.
If hot gas in both groups and clusters experienced the 
same level of heating, the luminosity of groups 
should be disproportionally lower since the 
heating in the adiabatic accretion shock is 
much greater in clusters, typically 
masking the amount of non-adiabatic heating required 
to account for the steep $L - T$ variation.

If gas and dark matter scale homologously 
within the virial radius and if all the gas that 
passes through the accretion shock remains within $r_{vir}$, 
the gas density should be the same at any fraction 
of $r_{vir}$ regardless of the mass of the cluster.
In this case the entropy factor $S = T/n_e^{2/3}$ 
is proportional to $T$.
Lloyd-Davies et al. (2000) have used the X-ray
data to estimate the entropy factor
$S_{0.1} = T/n_e^{2/3}$ at $0.1r_{vir}$ in group and cluster gas.
They find that when the 
$S_{0.1} \propto T$ relation is normalized to massive clusters, 
the values of $S_{0.1}$ for groups are too large to fit the 
same relation.  
Either some 
gas in groups flowed out because it was internally
heated or an insufficient amount of
gas entered the group at early times because the gas
was pre-heated before the accretion occurred.

This leads directly to the question we have addressed here:
Was the diffuse gas in galaxy groups and clusters pre-heated by 
some external process at high redshifts or was 
the heating internal, 
due to supernovae that accompanied normal star formation? 
Some recent studies of this question have invoked 
a universal cosmic pre-heating from sources 
unrelated to star formation. 
In view of its importance for galaxy formation, 
we have developed an ensemble of gas dynamical models
with and without pre-heating.

\subsection{External Pre-heating}

To investigate the consequences of universal pre-heating 
we study the accumulation of gas in group and 
cluster dark halos with three progressively more realistic 
assumptions: adiabatic AD flows with adiabatic shocks, 
CO flows with radiative cooling (Tozzi \& Norman 2000), 
and finally DO flows with mass dropout and radiative cooling 
(Knight \& Ponman 1997; Nulsen \& Fabian 1995, 1997).
In order to compare our results with those of previous 
studies, we also assume that the gas temperature and 
density are reset to uniform values at the moment 
when the pre-heating occurred, i.e. excess gas inside 
the dark halo is removed. 
For most of our models the reset is assumed to occur 
at time $t_h = 0.5$ Gyr or redshift $z_h = 9$.
In spite of its popularity in the current literature,
there is some uncertainty about the reset and 
pre-heating procedure. 
While very few stars have formed at these high redshifts 
and the only collapsed objects correspond to very small 
stellar systems, the tophat perturbations that we 
use to generate present day group and cluster halos 
have collected baryonic gas which is 
removed at time $t_h$ by the reset assumption.
We have also done many calculations with no reset.
The dark halos evolve in a normal way, initiated by 
a tophat perturbation and cooling toward an 
NFW profile in a flat $\Lambda$CDM universe. 
The final halo is insensitive to the tophat 
perturbation, i.e. the same halo 
can be produced by a variety of 
different tophat density amplitudes with appropriately 
chosen tophat radii.

In general we find that data in the $L - T$ plot 
(Figure 1) can be fit with AD, CO or DO flows 
with reset provided the pre-heating is sufficiently intense, 
1.3 to 3.9 keV/particle ($K_{34} = 0.25 - 0.76$).
However, these strongly pre-heated flows typically have 
entropies $S_{0.1}$ for groups and poor clusters that 
exceed observed values (Figure 2).
Values of the entropy and temperature that are 
mass-weighted within the virial radius 
obey the $S \propto T$ similarity relation except 
for the most strongly pre-heated group gas (Figure 4), 
but these cannot be compared with current observations.
However, a variety of computational and physical 
problems have appeared in our models with cosmic pre-heating. 
For example, weakly heated AD models 
are not only less successful in the $L - T$ plot, 
but also show irregularities due to transient shocks
following the reset that persist as 
irregularities in the entropy profiles $S(r)$ 
in the final AD1 models (Figure 3). 
$S_{0.1}$ may be increased by these spurious shocks.
With more computational effort these irregularities 
could probably be removed. 
But it is clear that adiabatic (AD) models cannot be correct 
in spite of their considerable success in fitting 
the $L - T$ data (with large pre-heating) 
and the entropy profiles (with little pre-heating): 
adiabatic flows cannot form stars 
nor can they radiate X-rays.

When radiative losses are included, as in the CO flows, 
we find that some  
pre-heated models agree with data the $L - T$ plot 
and have entropy profiles very similar 
to those of Tozzi \& Norman (2000) who make similar 
assumptions. 
However, group models with less pre-heating and all 
cluster models except COc4 have enormously massive and 
luminous X-ray singularities  
produced by emission from uncooled 
gas flowing into the center of the cluster.
Because of their luminous cores, 
these models have absurdly high X-ray luminosities and 
do not correspond to any object observed.
Another somewhat technical, but physically real,
aspect of CO flows is their tendency to produce 
non-linear thermally unstable drip waves that 
propagate into the core of the flow sometimes 
at high velocity. 
The drip waves are Rayleigh-Taylor unstable so 
they cannot remain perfectly coherent, 
unlike their representations in 
our one-dimensional models, and they will tend 
to be limited by mass dropout, not included in 
the CO models.
Finally, our results differ from those of Tozzi 
\& Norman (2000) in that the accretion shock 
is very close to the virial radius for all levels
of pre-heating considered.

To avoid these difficulties encountered in the 
CO models, we performed a final DO series of 
pre-heated models in which gas is allowed to 
cool to very low temperatures everywhere within 
the virial radius. 
This is a standard assumption used in models of 
cooling flows.
As before, 
strongly pre-heated DO models that fit the 
$L - T$ data have entropies $S_{0.1}$ 
that exceed observed values. 
While the ultra-luminous X-ray cores of the CO flows 
are no longer present in the DO models, 
the accumulated mass of cooled (dropout) mass 
can be large, particularly in massive clusters
where it exceeds the central mass densities 
of luminous elliptical galaxies by 10 - 100. 

We have also performed calculations of all types 
of externally pre-heated
flows with no uniform reset of the gas density and 
temperature at time $t_h$. 
In these models the cosmic gas temperature is 
heated by various amounts throughout the flow 
at $t_h$, but the density profile is not altered.
To reduce the final luminosity $L$ by the same 
amount as in the reset models, 
the no reset models require much higher levels 
of pre-heating to compensate for 
the energy radiated away from denser cores.
The COnr models are totally unacceptable 
because of the intense concentration of X-ray 
emission from the core where all the gas 
has cooled.
DOnr models are more similar to their 
DO counterparts since most of the heating 
is radiated away by dense gas in the core.
Although DOgnr flows are somewhat hotter
than the corresponding DOg flows,  
they nevertheless 
have lower central entropies $S_{0.1}$.

For a variety of reasons, none of our pre-heated models 
are particularly attractive, with or without reset.
In addition to gas dynamical concerns, the agency 
that heats the gas is unclear. 
Since the cosmic gas density is higher in the early 
universe, more energy is required to reach the 
same adiabat than at lower redshifts.
In the flat $\Lambda$CDM universe global 
pre-heating by Type II supernovae during star formation 
is unable to heat all the gas sufficiently 
at redshifts $z \gta 1$ to 
achieve agreement with galaxy group observations in 
the $L - T$ plane.
There may be enough energy in AGN to heat the gas 
at high redshifts, 
but it is unclear if this energy can be widely distributed 
in the intergalactic gas and not just concentrated near 
the AGN.

In comparing pre-heated models with 
observations, we place 
more emphasis on the $L - T$ plane  
than the $S_{0.1} - T$ plane.
The $S_{0.1}$ data are of lower quality because 
they typically refer to the outermost detectable 
X-ray emission from groups where the background corrections
may be troublesome.
In addition the virial radius must be estimated to 
derive $S_{0.1}$ from the observations.
Some of our models are also less certain 
in the inner parts of 
the flow $r \lta 0.1r_{vir}$ where the entropy 
can be increased by early shock waves or rapidly moving 
drip waves.

\subsection{Internal Heating by Supernovae}

In our final series of GA models the external pre-heating
assumption is replaced with internal heating 
by Type II supernovae associated with star formation at 
time $t_* = 2$ Gyrs and redshift 
$z_* = 3$ (see also Loewenstein 2000; Bryan 2000).
This internal heating is distributed within the 
accretion shock ($\sim r_{vir}$) in proportion 
to the local gas density. 
Immediately following $t_*$ a strong starburst wind
occurs in the GA group solutions.  
High entropy gas produced by the starburst shock 
currently extends beyond the virial radius.
Accompanying the SNII energy release is the formation 
of a luminous elliptical galaxy at the center of the group or 
cluster flow. 
As a guide, we have chosen the stellar mass of  
the NGC 2563 group to calibrate the amount of SNII energy 
delivered to the hot gas in groups. 
This same SNII heating per unit baryonic mass 
is applied to GA models for clusters.
We have explored the effect of varying the time $t_*$
of star formation and altering the distribution of the 
SNII energy within $r_{vir}$, but the final results are 
surprisingly insensitive to these parameters.
The GA models resemble DO models without reset 
since both include cooling mass dropout and have 
dense baryonic cores;  
the main difference is that the GA models are 
heated internally and DO models are heated universally.

While we have not attempted to produce a detailed 
model for the NGC 2563 group, we find that 
the X-ray luminosity and temperature observed in 
NGC 2563 are almost exactly matched by our GA group model 
at time $t_n = 13$ Gyrs when heated 
with SNII energy expected from a Salpeter IMF. 
The data in the $S_{0.1} - T$ plane is also nicely 
fit by the group GA models.
Finally, we find that our standard GA group models
agree well with the entropy distribution
$S(r)$ observed in NGC 2563.
Excellent agreement with observations 
in both the $L - T$ and 
$S - T$ plots can be achieved with 
a range of 8 in the SNII energy released 
per unit stellar mass, corresponding to 
different IMFs.
However, for the best results we require that 
at least $\sim 0.2$ 
of the energy released by SNII from early star formation
directly heats the hot gas. 
The local environment of the early SNII explosions 
may differ from star-forming regions in our Galaxy 
in that the supernova blast waves propagate into 
the hot gas, not into 
cold, dense clouds which can radiate away much of 
the supernova energy.
Of course the hot gas heating efficiency could be 
$< 0.2$ if the IMF were more top-heavy than 
expected.
We also present GA models for clusters, although
our physical model of monolithic growth (without
detailed hierarchical merging) is less appropriate
for cluster masses.

Mushotzky and Scharf (1997) have 
shown that the evolution of data in 
the $L - T$ plane is very weak out to redshifts 
$z \sim 0.4$.
Our GA group models also exhibit this same slow 
evolution.
For example, 
the mean bolometric X-ray luminosity and gas temperature
for the four GA models we consider differ 
at $z = 1$ by only 
45 and -7 percent respectively from the same 
averages at zero redshift, 
indicating that the temperature is stable and that 
the luminosity may be declining slightly. 

In view of the generally good agreement with observations, 
we conclude that galaxy group models 
with internal SNII heating fit the data better 
and more plausibly than 
any of the externally pre-heated models we 
have considered. 
The production of SNII energy and the transfer of 
this energy to the hot gas may be a concern 
for the internal heating models we prefer, 
but there has been no satisfactory explanation of  
the sources of the much greater 
energy associated with universal, external pre-heating 
at high redshifts.

%\end{document}

\acknowledgments
We wish to thank referee Paolo Tozzi for 
helpful criticism and advice.
Studies of the evolution of hot gas in elliptical galaxies
at UC Santa Cruz are supported by
NASA grant NAG 5-3060 and NSF grant
AST-9802994 for which we are very grateful.

%\clearpage

\vskip.1in
\figcaption[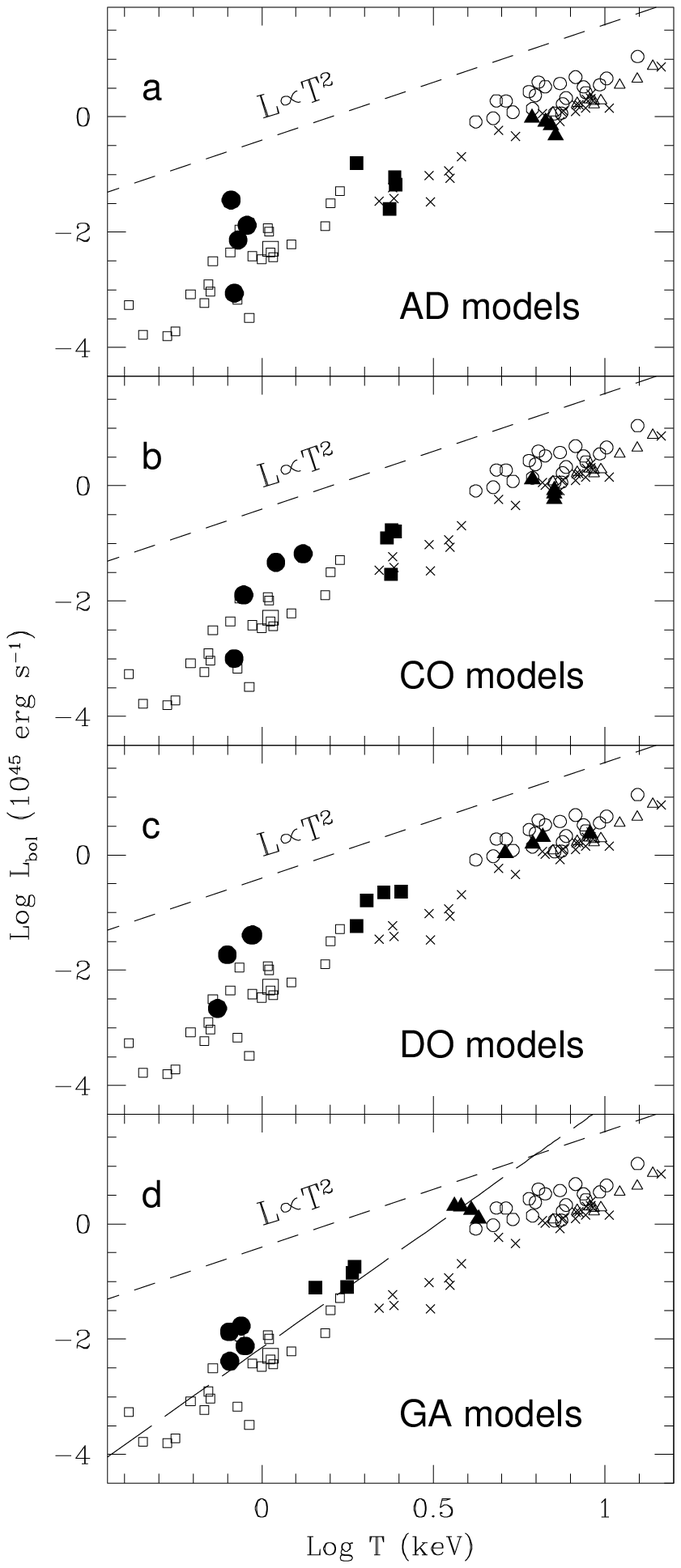]{
Variation of X-ray bolometric luminosity with emission 
weighted temperature. 
Each panel shows the data:
{\it open squares:} groups from Helsdon \& Ponman (2000);
{\it crosses:} clusters from Arnaud \& Evrard (1999); 
{\it open circles and triangles:} clusters from 
Allen \& Fabian (1998), the open triangles refer to 
non-cooling flow clusters.
The dashed line shows the (un-normalized) 
$L \propto T^2$ relation 
expected if the hot gas were adiabatic and emitted 
bremsstrahlung radiation.
The four panels a $\rightarrow$ d show results 
of calculations for adiabatic (AD), radiative cooling
(CO), mass dropout (DO), and galaxy formation (GA) 
models.
The filled circles, squares and triangles are 
results of gas dynamical calculations at time 13 Gyrs 
for groups, poor clusters and clusters respectively.
Note that the computed luminosities and 
temperatures for the two DO models for groups DOg1 and DOg2  
are almost identical. 
\label{fig1}}

\vskip.1in
\figcaption[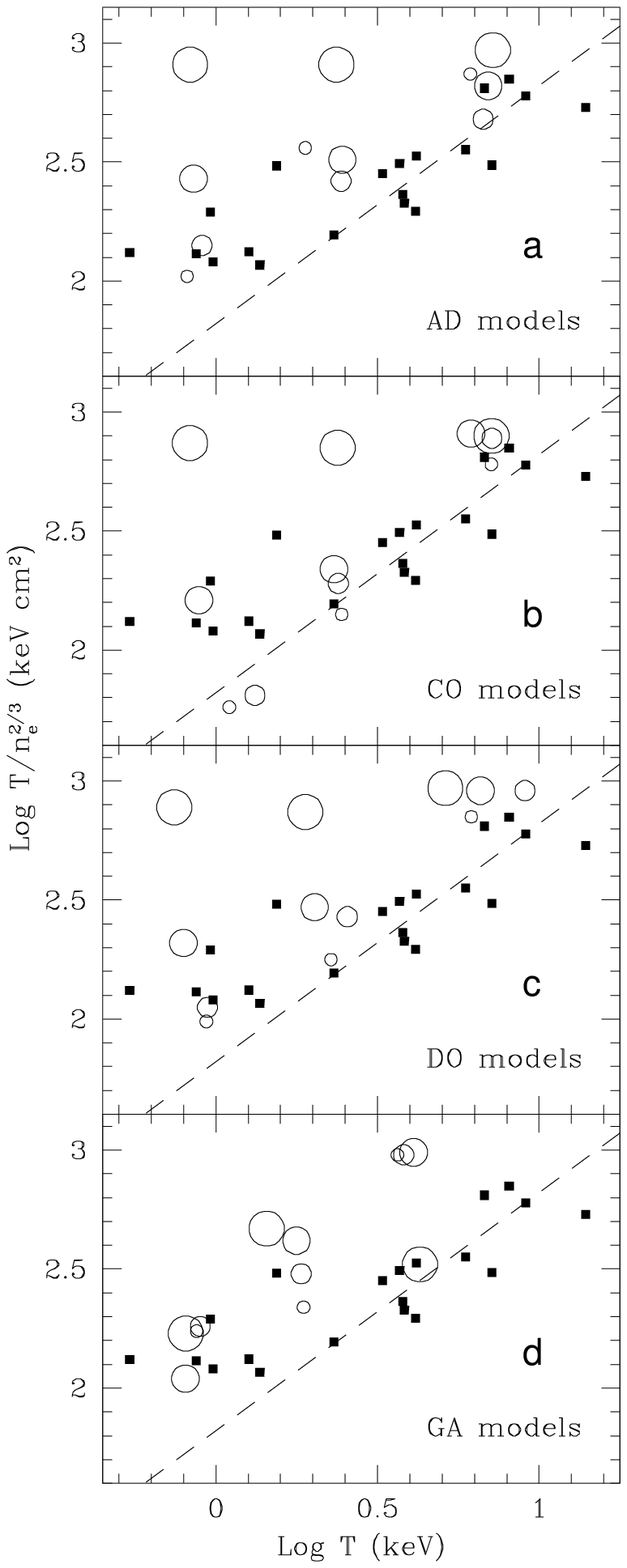]{
Plot of the entropy factor $S = T n_e^{-2/3}$ evaluated 
at radius $0.1r_{vir}$ against emission-weighted 
temperature.
Each panel shows the entropy data (filled squares) from 
Lloyd-Davies, Ponman, \& Cannon (2000) with a 
dashed line $S \propto T$ passing through the 
rich clusters.
Open circles in each of 
the four panels a $\rightarrow$ d show results
of calculations for adiabatic (AD), radiative cooling
(CO), mass dropout (DO), and galaxy formation (GA)
models.
The radius of the open circle increases with the 
amount of heating as characterized by the nomenclature 
$1 \rightarrow 4$ as described in the text. 
\label{fig2}}

\vskip.1in
\figcaption[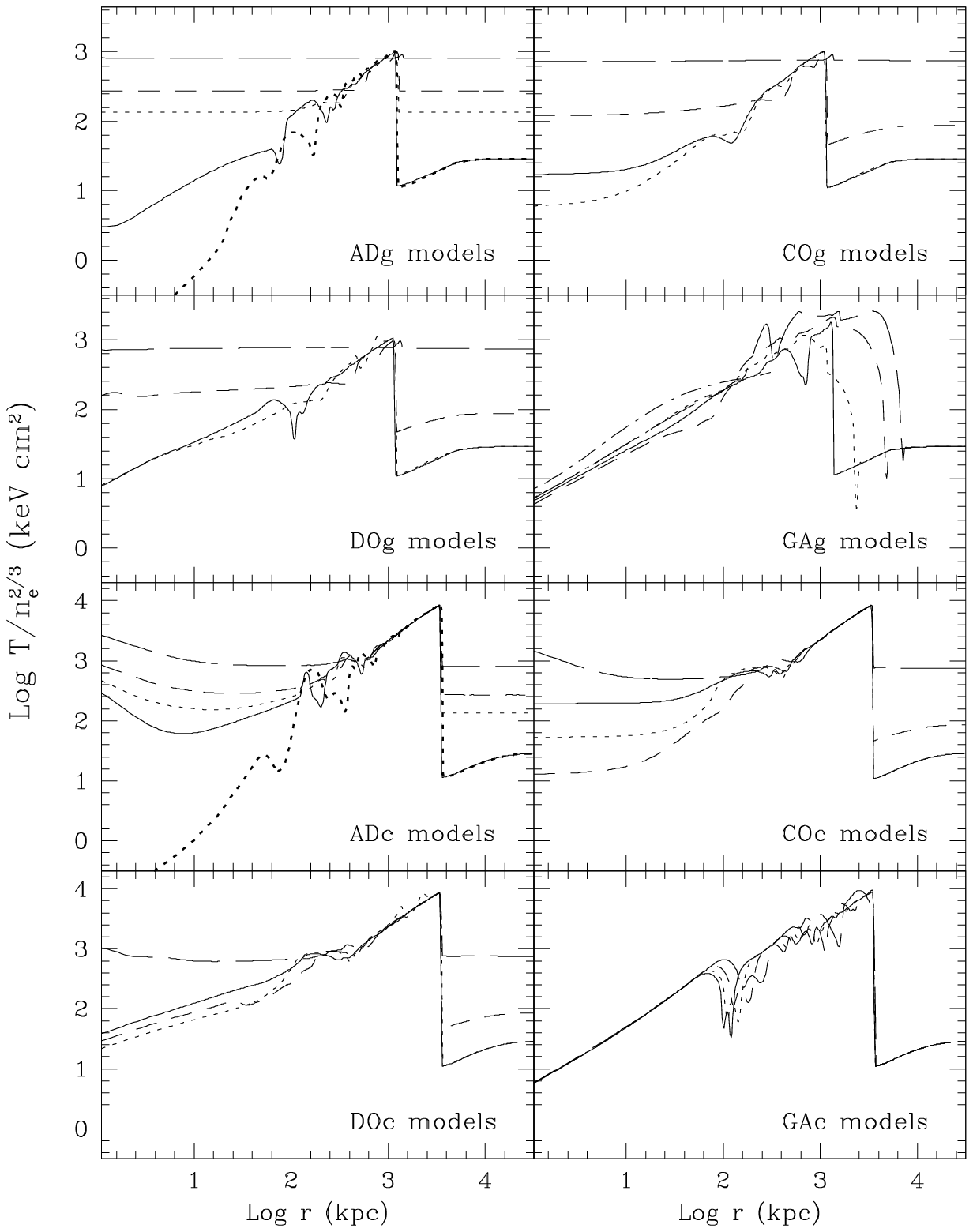]{
Profiles of the entropy factor $S(r) = T n_e^{-2/3}$ 
for adiabatic (AD), radiative cooling
(CO), mass dropout (DO), and galaxy formation (GA)
models.
The ``g'' or ``c'' refer to models 
of groups and clusters respectively.
The amount of heating described 
in the text with the nomenclature 
$1 \rightarrow 4$ is represented for each model with 
solid, dotted, short-dashed and long-dashed lines
respectively.
The heavy dotted lines for ADg  and ADc models refer 
to calculated profiles without reset.
The dot-dashed line in the GAg panel shows the 
observed entropy profile for group NGC 2563 
from Trinchieri, Fabbiano, \& Kim (1997).
\label{fig3}}

\vskip.1in
\figcaption[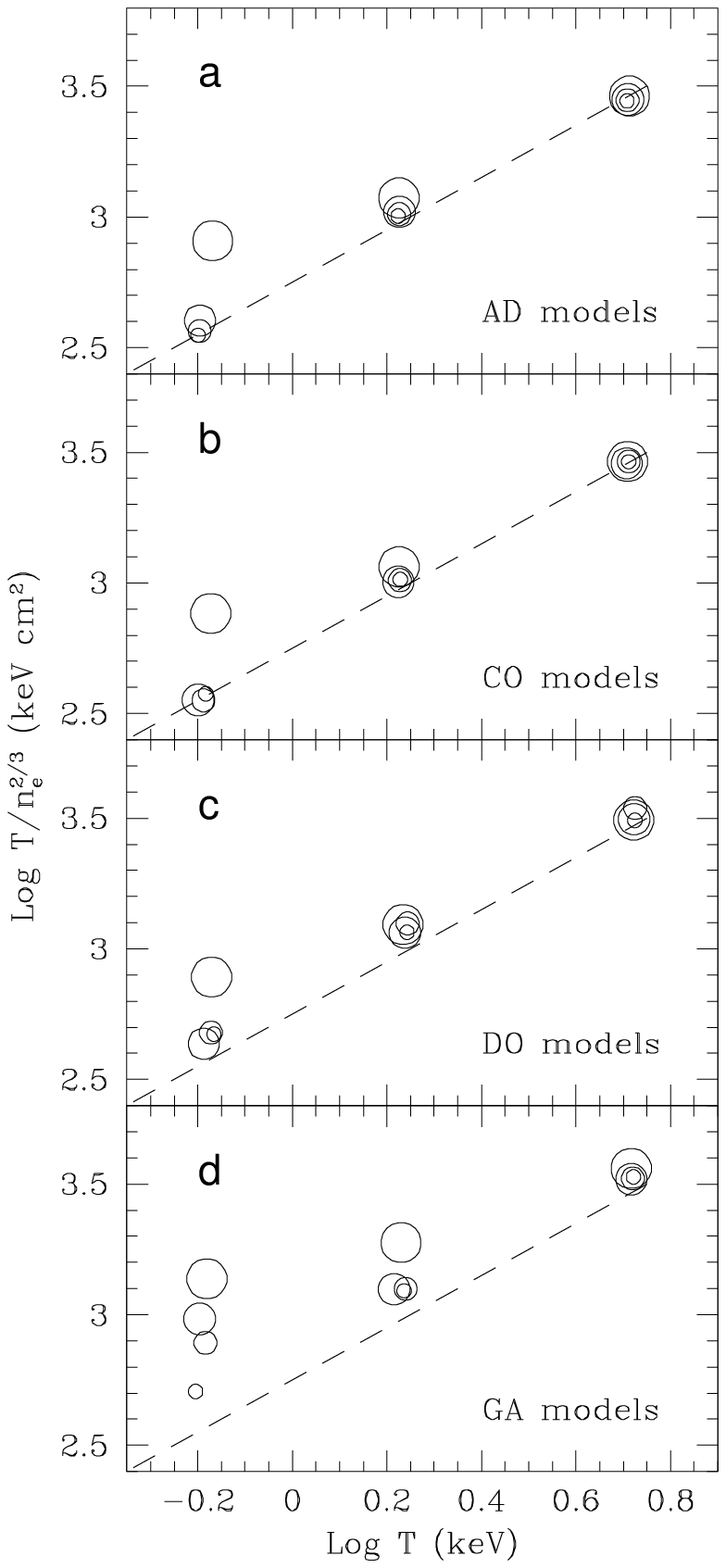]{
Plot of mass-weighted values of the entropy 
factor $S(r) = T n_e^{-2/3}$ and temperature 
both averaged within the virial radius.
The four panels a $\rightarrow$ d show results
of calculations for adiabatic (AD), radiative cooling
(CO), mass dropout (DO), and galaxy formation (GA)
models.
Results in each panel are shown with clusters
of open circles for groups, poor clusters and
clusters, respectively from left to right.
The radius of the open circle increases with the
amount of heating as characterized by the nomenclature
$1 \rightarrow 4$ as described in the text.
The dashed line shows the slope $S \propto T$ 
expected for adiabatic hot gas configurations.
\label{fig4}}

\vskip.1in
\figcaption[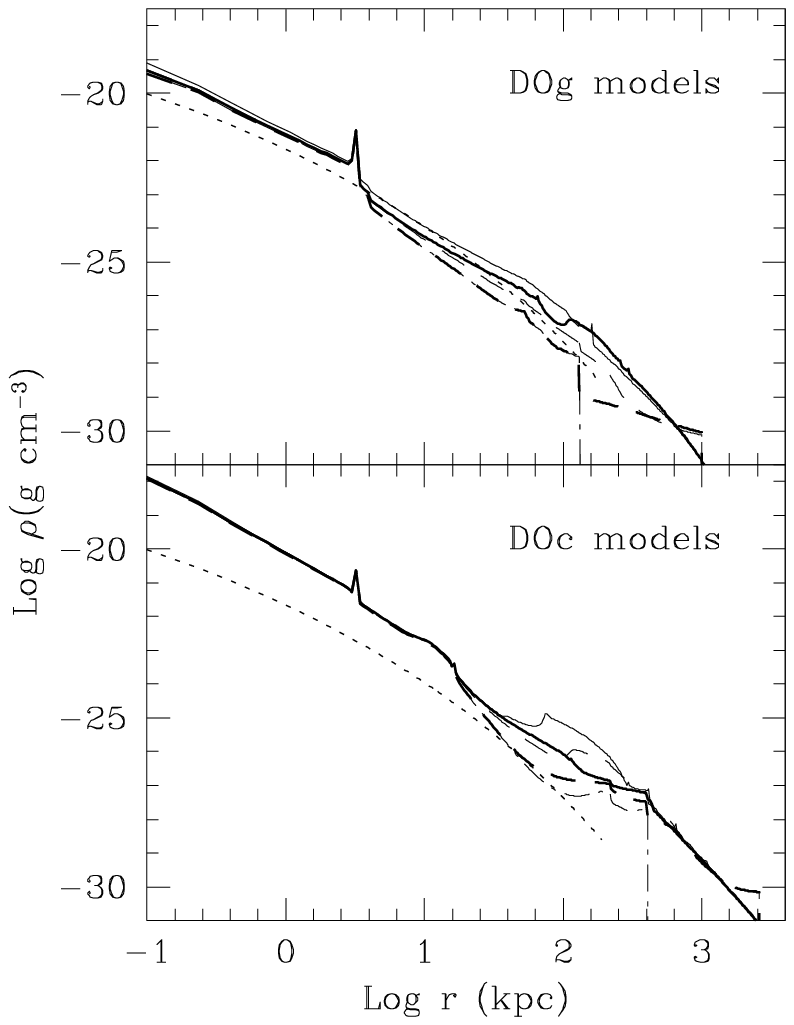]{
Density profiles of cooled (dropout) gas at 
13 Gyrs for groups (upper panel) and clusters 
(lower panel).
{\it Heavy solid lines:} groups and clusters 
with no pre-heating, DOg1 and DOc1;
{\it Heavy dashed lines:} groups and clusters 
with maximum pre-heating, DOg4 and DOc4;
{\it Light solid lines:} groups and clusters 
with no pre-heating and no reset, DOg1nr and DOc1nr;
{\it Light dashed lines:} groups and clusters 
with maximum pre-heating and no reset, DOg4nr and DOc4nr;
{\it Light dot-dashed lines:} groups and clusters
at time 0.5 Gyr, just before reset;
{\it Dotted lines:} de Vaucouleurs mass profile 
for elliptical galaxy NGC 2563.
\label{fig5}}

\vskip.1in
\figcaption[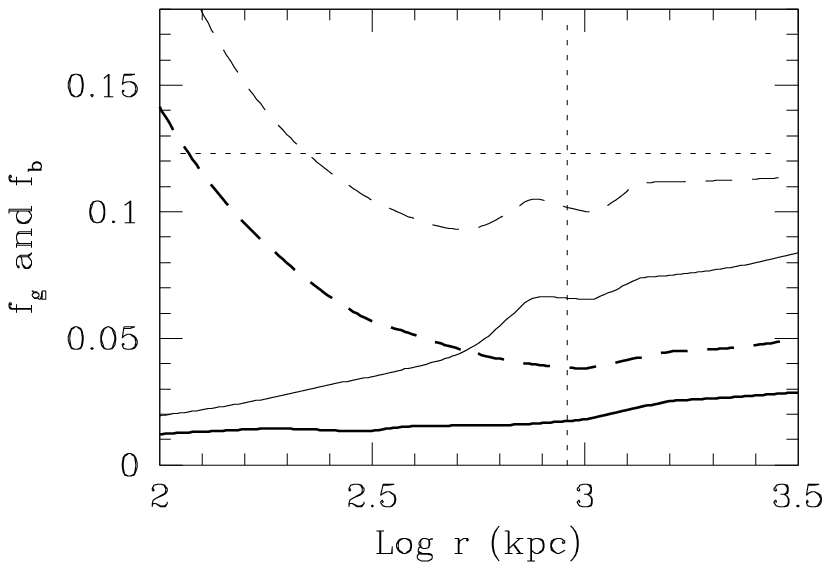]{
Baryon fraction $f_b$ (dashed lines) 
and gas fraction $f_g$ (solid lines) at 13 
Gyrs for GAg1 models (light lines) 
and GAg4 models (heavy lines).
The cosmic baryon fraction is shown with a 
horizontal dashed line.
The virial radius $r_{vir}$ is shown with 
a vertical dashed line.
\label{fig6}}

\clearpage

\begin{deluxetable}{lcccc}
\tabletypesize{\footnotesize}
\normalsize
%\scriptsize
%\tabletypesize{\footnotesize}
%\rotate
%\tablewidth{30pc}
\tablewidth{10cm}
\tablenum{1}
\tablecolumns{5}
\tablecaption{ADIABATIC MODELS\tablenotemark{a}}
\tablehead{
\colhead{Model} &
\colhead{$\langle T \rangle$\tablenotemark{b}} &
\colhead{$\log L$\tablenotemark{c}} &
\colhead{$\log S(0.1r_{vir})$\tablenotemark{d}} &
\colhead{$f_B(r_{vir})$\tablenotemark{e}} \cr
\colhead{} &
\colhead{(keV)} &
\colhead{(ergs/s)} &
\colhead{(keV cm$^2$)} &
\colhead{} \cr
}
\startdata
ADg1 &     0 .814 & 43.55   &   2.02  & 0.0918 \cr
ADg2 &     0 .906 & 43.12   &   2.15  & 0.0900 \cr
ADg3 &     0 .853 & 42.86   &   2.43  & 0.0856 \cr
ADg4 &     0 .843 & 41.94   &   2.91  & 0.0420 \cr
\cr
ADpc1 &    1.89  &  44.19   & 2.55  & 0.0931 \cr
ADpc2 &    2.44  &  43.95   & 2.42  & 0.0916 \cr
ADpc3 &    2.46  &  43.83   & 2.51  & 0.0899 \cr
ADpc4 &    2.36  &  43.41   & 2.91  & 0.0811 \cr
\cr
ADc1  &     6.13 &  44.99 & 2.87 &  0.0907 \cr
ADc2  &     6.72 &  44.91 & 2.68 &  0.0902 \cr
ADc3  &     6.96 &  44.85 & 2.82 &  0.0897 \cr
ADc4  &     7.20 &  44.68 & 2.97 &  0.0877 \cr
\cr
ADg1nr &     0 .392 & 46.14   &   1.83  & 0.121 \cr
ADg4nr &     0 .425 & 44.76   &   1.85  & 0.068 \cr
ADpc1nr &    0.760  & 46.41   &   1.63  & 0.121 \cr
ADpc4nr &    0.908 &  45.47   &   2.22  & 0.100 \cr
ADc1nr  &    1.714 &  46.78   &   2.44 &  0.117 \cr
ADc4nr  &    2.399 &  46.14   &   2.81 &  0.114 \cr

\enddata
\footnotesize
\tablenotetext{a}{Nomenclature: g, pc, and c represent group, 
poor cluster and cluster potentials; 1 indicates no 
pre-heating and 3 $\rightarrow$ 4 represent 
increasing levels of pre-heating; nr indicates no reset.}
\tablenotetext{b}{Emission weighted temperature within
$r_{vir} = 0.88$, 1.47, and 2.63 Mpc for group, poor cluster and
cluster}
\tablenotetext{c}{Bolometric X-ray luminosity within $r_{vir}$}
\tablenotetext{d}{Entropy factor at $0.1r_{vir}$}
\tablenotetext{e}{Baryon fraction in hot gas within $r_{vir}$;
the total baryon fraction including dropout mass is shown in
parenthesis}
\end{deluxetable}
\normalsize

\clearpage

\begin{deluxetable}{lclccc}
\tabletypesize{\footnotesize}
\normalsize
%\scriptsize
%\tabletypesize{\footnotesize}
%\rotate
%\tablewidth{30pc}
\tablewidth{12cm}
\tablenum{2}
\tablecolumns{6}
\tablecaption{COOLING MODELS\tablenotemark{a}}
\tablehead{
\colhead{Model} &
\colhead{$\langle T \rangle$\tablenotemark{b}} &
\colhead{$\log L$\tablenotemark{c}} &
\colhead{$\log S(0.1r_{vir})$\tablenotemark{d}} &
\colhead{$f_B(r_{vir})$\tablenotemark{e}} &
\colhead{$\log M_{cold}$\tablenotemark{f}} \cr
\colhead{} &
\colhead{(keV)} &
\colhead{(ergs/s)} &
\colhead{(keV cm$^2$)} &
\colhead{} &
\colhead{($M_{\odot}$)} \cr
}
\startdata
COg1 & 1.10 & 43.67(44.55) & 1.76 & 0.083(0.095) & 11.77 \cr
COg2 & 1.32 & 43.82(44.62) & 1.81 & 0.080(0.096) & 11.59 \cr
COg3 & 0.886 & 43.11 & 2.21 & 0.092(0.092) & \nodata \cr
COg4 & 0.831 & 42.00 & 2.87 & 0.045(0.045) & \nodata \cr
\cr
COpc1 & 2.45 & 44.21(44.80) & 2.15 & 0.09(0.094) & 11.94 \cr
COpc2 & 2.39 & 44.24(44.90) & 2.28 & 0.091(0.094) & 11.80 \cr
COpc3 & 2.32 & 44.10 & 2.34 & 0.092(0.092) & \nodata \cr
COpc4 & 2.38 & 43.47 & 2.85 & 0.082(0.082) & \nodata \cr
\cr
COc1 & 7.12 & 44.87(45.40) & 2.78 & 0.088(0.091) & 12.55 \cr
COc2 & 7.14 & 44.92(46.04) & 2.89 & 0.088(0.091) & 12.53 \cr
COc3 & 6.15 & 45.11(45.90) & 2.91 & 0.089(0.090) & 12.16 \cr
COc4 & 7.12 & 44.77 & 2.90 & 0.088(0.088) & \nodata \cr
\cr
COg1nr & 2.58  & 43.00(45.35) & 2.40 & 0.065(0.126) & 12.48 \cr
COg4nr & 1.28  & 42.28(44.10) & 2.53 & 0.047(0.079) & 12.16 \cr
COc1nr & 14.93 & 44.91(47.69) & 2.94 & 0.085(0.118) & 13.60 \cr
COc4nr & 8.54  & 44.87(46.93) & 2.97 & 0.092(0.114) & 13.41 \cr
\enddata
\footnotesize
\tablenotetext{a}{Nomenclature: g, pc, and c represent group,
poor cluster and cluster potentials; 1 indicates no
pre-heating and 3 $\rightarrow$ 4 represent
increasing levels of pre-heating; nr indicates no reset.}
\tablenotetext{b}{Emission weighted temperature within 
$r_{vir} = 0.88$, 1.47, and 2.63 Mpc for group, poor cluster and
cluster}
\tablenotetext{c}{Bolometric X-ray luminosity within $r_{vir}$;
values of $\log(L + L_{core})$ are shown in 
parenthesis when gas cools in the central zone.}
\tablenotetext{d}{Entropy factor at $0.1r_{vir}$}
\tablenotetext{e}{Baryon fraction in hot gas within $r_{vir}$;
the total baryon fraction including dropout mass is shown  
in parenthesis}
\tablenotetext{f}{Mass cooled at $r = 0$}
\end{deluxetable}
\normalsize

\clearpage

\begin{deluxetable}{lccccc}
\tabletypesize{\footnotesize}
\normalsize
%\scriptsize
%\tabletypesize{\footnotesize}
%\rotate
%\tablewidth{30pc}
\tablewidth{12cm}
\tablenum{3}
\tablecolumns{6}
\tablecaption{MASS DROPOUT MODELS\tablenotemark{a}}
\tablehead{
\colhead{Model} &
\colhead{$\langle T \rangle$\tablenotemark{b}} &
\colhead{$\log L$\tablenotemark{c}} &
\colhead{$\log S(0.1r_{vir})$\tablenotemark{d}} &
\colhead{$f_B(r_{vir})$\tablenotemark{e}} &
\colhead{$\log M_{cold}$\tablenotemark{f}} \cr
\colhead{} &
\colhead{(keV)} &
\colhead{(ergs/s)} &
\colhead{(keV cm$^2$)} &
\colhead{} &
\colhead{($M_{\odot}$)} \cr
}
\startdata
DOg1 & 0.933 & 43.61 & 1.99 & 0.072(0.114) & 12.31 \cr
DOg2 & 0.941 & 43.61 & 2.05 & 0.072(0.115) & 12.35 \cr
DOg3 & 0.795 & 43.27 & 2.32 & 0.081(0.111) & 12.17 \cr
DOg4 & 0.743 & 42.34 & 2.89 & 0.045(0.067) & 11.98 \cr
\cr
DOpc1 & 2.27 & 44.35 & 2.25 & 0.084(0.106) & 12.68 \cr
DOpc2 & 2.55 & 44.36 & 2.43 & 0.077(0.107) & 12.81 \cr
DOpc3 & 2.02 & 44.21 & 2.47 & 0.084(0.105) & 12.64 \cr
DOpc4 & 1.89 & 43.76 & 2.87 & 0.080(0.095) & 12.48 \cr
\cr
DOc1 & 6.17 & 45.20 & 2.85 & 0.084(0.100) & 13.27 \cr
DOc2 & 9.05 & 45.36 & 2.96 & 0.076(0.102) & 13.47 \cr
DOc3 & 6.59 & 45.31 & 2.96 & 0.084(0.100) & 13.29 \cr
DOc4 & 5.13 & 45.03 & 2.97 & 0.085(0.098) & 13.17 \cr
\cr
DOg1nr & 1.24 & 43.59 & 1.98 & 0.066(0.123) & 12.45 \cr
DOg2nr & 1.54 & 43.89 & 1.95 & 0.069(0.129) & 12.48 \cr
DOg3nr & 1.29 & 43.49 & 2.24 & 0.083(0.125) & 12.33 \cr
DOg4nr & 1.43 & 43.31 & 2.25 & 0.050(0.079) & 12.12 \cr
\cr
DOc1nr & 6.74 & 45.47 & 2.86 & 0.088(0.117) & 13.55 \cr
DOc2nr & 6.39 & 45.45 & 2.84 & 0.080(0.119) & 13.67 \cr
DOc3nr & 4.77 & 45.70 & 2.79 & 0.088(0.118) & 13.55 \cr
DOc4nr & 5.57 & 45.57 & 2.76 & 0.095(0.115) & 13.37 \cr
\enddata
\footnotesize
\tablenotetext{a}{Nomenclature: g, pc, and c represent group,
poor cluster and cluster potentials; 1 indicates no
pre-heating and 3 $\rightarrow$ 4 represent
increasing levels of pre-heating; nr indicates no reset.}
\tablenotetext{b}{Emission weighted temperature within
$r_{vir} = 0.88$, 1.47, and 2.63 Mpc for group, poor cluster and
cluster}
\tablenotetext{c}{Bolometric X-ray luminosity within $r_{vir}$}
\tablenotetext{d}{Entropy factor at $0.1r_{vir}$}
\tablenotetext{e}{Baryon fraction in hot gas within $r_{vir}$;
the total baryon fraction including dropout mass is shown in
parenthesis}
\tablenotetext{f}{Total cooled dropout mass}
\end{deluxetable}
\normalsize

\clearpage

\begin{deluxetable}{lccccc}
\tabletypesize{\footnotesize}
\normalsize
%\scriptsize
%\tabletypesize{\footnotesize}
%\rotate
%\tablewidth{30pc}
\tablewidth{12cm}
\tablenum{4}
\tablecolumns{6}
\tablecaption{GALAXY FORMATION MODELS\tablenotemark{a}}
\tablehead{
\colhead{Model} &
\colhead{$\langle T \rangle$\tablenotemark{b}} &
\colhead{$\log L$\tablenotemark{c}} &
\colhead{$\log S(0.1r_{vir})$\tablenotemark{d}} &
\colhead{$f_B(r_{vir})$\tablenotemark{e}} &
\colhead{$\log M_{cold}$\tablenotemark{f}} \cr
\colhead{} &
\colhead{(keV)} &
\colhead{(ergs/s)} &
\colhead{(keV cm$^2$)} &
\colhead{} &
\colhead{($M_{\odot}$)} \cr
}
\startdata
GAg1 & 0.871 & 43.23 & 2.24 & 0.066(0.102) & 11.97 \cr
GAg2 & 0.894 & 42.88 & 2.26 & 0.040(0.071) & 11.81 \cr
GAg3 & 0.804 & 43.13 & 2.04 & 0.027(0.052) & 11.58 \cr
GAg4 & 0.807 & 42.62 & 2.23 & 0.017(0.039) & 11.22 \cr
\cr
GApc1 & 1.866 & 44.26 & 2.34 & 0.087(0.123) & 12.86 \cr
GApc2 & 1.835 & 44.15 & 2.48 & 0.088(0.123) & 12.84 \cr
GApc3 & 1.774 & 43.91 & 2.62 & 0.083(0.116) & 12.81 \cr
GApc4 & 1.435 & 43.89 & 2.67 & 0.054(0.083) & 12.74 \cr
\cr
GAc1 & 3.644 & 45.31 & 2.98 & 0.086(0.118) & 13.57 \cr
GAc2 & 3.812 & 45.30 & 2.98 & 0.087(0.118) & 13.56 \cr
GAc3 & 4.091 & 45.25 & 2.99 & 0.089(0.118) & 13.54 \cr
GAc4 & 4.281 & 45.08 & 2.52 & 0.085(0.113) & 13.51 \cr
\enddata
\footnotesize
\tablenotetext{a}{Nomenclature: g, pc, and c represent group,
poor cluster and cluster potentials; 
1, 2, 3 and 4 represent increasing supernova heating,
$\eta^* = 0.5$, 1, 2, and 4 respectively} 
\tablenotetext{b}{Emission weighted temperature within
$r_{vir} = 0.88$, 1.47, and 2.63 Mpc for group, poor cluster and
cluster}
\tablenotetext{c}{Bolometric X-ray luminosity within $r_{vir}$}
\tablenotetext{d}{Entropy factor at $0.1r_{vir}$}
\tablenotetext{e}{Baryon fraction in hot gas within $r_{vir}$;
the total baryon fraction including dropout mass is shown in
parenthesis}
\tablenotetext{f}{Total cooled dropout mass within $r_{vir}$}
\end{deluxetable}
\normalsize

\end{document}